\documentclass[%
 reprint,
 amsmath,amssymb,
 aps,
 pra,
 floatfix,
 nofootinbib,
]{revtex4-2}

\usepackage{graphicx}
\usepackage{dcolumn}
\usepackage{bm}
\usepackage{gensymb}

\begin{document}

\preprint{APS/123-QED}

\title{Accurate determination of an alkali--inert gas diffusion coefficient using coherent transient emission from a density grating}

\author{A. Pouliot}
\email{alexpouliot@live.com}
 \affiliation{Department of Physics and Astronomy, York University, 4700 Keele St. Toronto ON Canada, M3J 1P3.}
\author{G. Carlse}%
\affiliation{Department of Physics and Astronomy, York University, 4700 Keele St. Toronto ON Canada, M3J 1P3.}
\author{H. C. Beica}%
\affiliation{Department of Physics and Astronomy, York University, 4700 Keele St. Toronto ON Canada, M3J 1P3.}
\author{T. Vacheresse}%
\affiliation{Department of Physics and Astronomy, York University, 4700 Keele St. Toronto ON Canada, M3J 1P3.}
\author{A. Kumarakrishnan}%
\email{akumar@york.ca}
\affiliation{Department of Physics and Astronomy, York University, 4700 Keele St. Toronto ON Canada, M3J 1P3.}


\author{U. Shim}
\altaffiliation[Current address: ]{ONTO Innovation Inc. 59-2 Sukwoo--dong Hwasung City, Gyeonggi--Do, South Korea 445-170}
\affiliation{
Department of Physics, New York University
}
\author{S. B. Cahn}
\altaffiliation[Current address: ]{Physics Department, Yale University}
\affiliation{
Department of Physics, New York University
}
\author{A. Turlapov}
\altaffiliation[Current address: ]{Institute of Applied Physics, Russian Academy of Sciences, Nizhniy Novgorod, 603000, Russia}
\affiliation{
Department of Physics, New York University
}
\author{T. Sleator}
\affiliation{
Department of Physics, New York University
}


\date{\today}

\begin{abstract}
We demonstrate a new technique for the accurate measurement of diffusion coefficients for alkali vapor in an inert buffer gas. The measurement was performed by establishing a spatially periodic density grating in isotopically pure $^{87}$Rb vapor and observing the decaying coherent emission from the grating due to the diffusive motion of the vapor through N$_2$ buffer gas. We obtain a diffusion coefficient of $0.245 \pm0.002 ~\textrm{cm}^{2}/\textrm{s}$ at 50$\degree$C and 564~Torr. Scaling to atmospheric pressure, we obtain $D_0 = 0.1819 \pm 0.0024~\textrm{cm}^2/\textrm{s}$. To the best of our knowledge, this represents the most accurate determination of the Rb--N$_2$ diffusion coefficient to date. Our measurements can be extended to different buffer gases and alkali vapors used for magnetometry and can be used to constrain theoretical diffusion models for these systems.
\begin{description}
\item[PACS Numbers]
\end{description}
\end{abstract}

\maketitle


\section{\label{sec:intro}Introduction}

During the last forty years, there have been significant improvements in the sensitivity of vapor cell magnetometers used for the detection of small magnetic fields and magnetic anomalies. The development of spin--exchange relaxation–-free (SERF) atomic magnetometers \cite{Allred} has allowed these sensors to reach sensitivities below $1~\textrm{fT}/\textrm{Hz}^{1/2}$, competing with, and often surpassing, superconducting (SQUID) magnetometers \cite{SQUID} to be the most precise magnetometers in the world \cite{SERFReview}. Atomic magnetometers operate by optically pumping alkali vapor into a specific internal state, thereby aligning the individual magnetic dipole moments of atoms. This net magnetic moment will oscillate at the Larmor frequency, which is uniquely determined by the external magnetic field. In a conventional, time--domain magnetometer, the Larmor frequency is measured by observing the absorption of a weak probe laser \cite{Mlynek}. SERF magnetometers achieve high precision by preserving the alignment over extended time scales. This is achieved by using high alkali densities and specific concentrations of buffer, and quenching gases. Under these conditions the optically--pumped alkali vapor slowly diffuses with minimal decoherence due to radiation trapping and spin--disrupting collisions. 

To optimize these devices, it is necessary to develop detailed models of optical pumping of the D1 and D2 lines in alkali atoms \cite{bermanDensityMatrix,brynle,spiedefense}, and make precise measurements of collisional cross sections \cite{HapperOptPump,Baranga,Jenkins,Speller,Wagshul,allardkielkopf} and diffusion coefficients \cite{Happer,Ishikawa} for relevant alkali and inert gas mixtures. Other motivations for precise diffusion measurements include spin--polarized, high--resolution imaging using noble gases \cite{HapperPolImg}, and mesospheric magnetometry involving sodium vapor for the monitoring of the Earth's ocean currents and interior dynamics \cite{budkerSodium}. 

Previous measurements of diffusion coefficients have involved analyzing transient signals associated with the optical pumping of alkali vapors \cite{McNeal,Franz,Happer,Wagshul,EricksonThesis,Parniak}, measuring the amplitude decay of spin echoes in a magnetic gradient \cite{Ishikawa,Ishikawa2}, and analyzing the spectrum of transmitted probe light well below the shot--noise limit to directly observe atomic motion \cite{Aoki}. The diffusion coefficient may also be determined indirectly via measurements of collision cross--sections \cite{bloembergen}. Diffusion coefficients can be inferred from these cross--sections, however such an approach would rely on the accuracy of the intermolecular potentials used by the Chapman--Enskog formalism \cite{Chapman,cussler2009}. 

Table~\ref{tab:diff} summarizes representative values of the \mbox{Rb--N$_2$} diffusion coefficient. The smallest uncertainty prior to this work was achieved by reference \cite{Ishikawa} (2.5\%). The discrepancy between measurements utilising different techniques emphasizes the necessity of a variety of methods, subject to different systematic effects, in arriving at a more reliable measurement. Additionally, accurate measurements of the diffusion coefficient constrain theoretical models of many particle systems. 

\begin{table*}
\caption{\label{tab:diff}%
Representative measurements of the $\textrm{Rb}-\textrm{N}_2$ diffusion coefficient at atmospheric pressure $D_0$. Uncertainties are provided where available.
}
\begin{ruledtabular}
\begin{tabular}{llll}
\textrm{Reference}&
\textrm{Technique}&
\textrm{$D_0$ ($\textrm{cm}^2/\textrm{s}$)}&
\textrm{$D_0$ rescaled to 50$\degree$C\footnote{Rescaled using $D\propto T^{3/2}$ \cite{Chapman,cussler2009}} ($\textrm{cm}^2/\textrm{s}$)}\\
\colrule
Wagshul et al. \cite{Wagshul} & Optical pumping relaxation & 0.28 at $150\degree\textrm{C}$\footnote{Average of values taken at various pressures and rescaled to 1 atm} & 0.19\\
Zeng et al. \cite{Happer} & Optical pumping relaxation & 0.20 at $70\degree C$ & 0.18\\
Ishikawa et al.  \cite{Ishikawa} & Magnetic resonance echo & $0.159\pm0.004$ at $60\degree \textrm{C}$ & $0.152\pm0.004$\\
McNeal et al. \cite{McNeal} & Optical pumping relaxation & 0.33 at 55$\degree\textrm{C}$ & 0.32\\
Franz et al. \cite{Franz} & Optical pumping relaxation & 0.16 at $32\degree\textrm{C}$ & 0.18\\
Erickson \cite{EricksonThesis} & Optical pumping relaxation & 0.30 at $180\degree\textrm{C}$ & 0.18\\
This work & Dephasing of density grating & $0.1819\pm0.0024$ at $50\degree\textrm{C}$ & $0.1819\pm0.0024$\\
\end{tabular}
\end{ruledtabular}
\end{table*}

In this paper, we present a contrasting technique that directly measures diffusion from the decay time of a long--lived, coherent transient signal with a simple functional form. The signal arises from a density grating which is insensitive to magnetic fields and magnetic field gradients. The timescale of the decay also shows a characteristic dependence on the grating spacing, which can be varied and measured precisely. The characteristic dependence also provides a good systematic check of the accuracy. As a result, this technique appears to be suitable for accurate and precise measurements of diffusion coefficients.

The rest of the paper is organized as follows: In Sec.~\ref{sec:theory} we contrast a traditional, time--domain magnetometer with one based on spatially--periodic atomic coherences. We demonstrate that for specific excitation polarizations and energy level schemes, it is also possible to realize density gratings with the same periodicity that are insensitive to magnetic fields. We explain how the characteristic decay times of these gratings can be exploited for measurements of diffusion coefficients. In Sec.~\ref{sec:Experiment} we describe the experimental details. The diffusion coefficient measurement is presented in Sec.~\ref{sec:results}. We conclude with a discussion of the impact of this work on the development of magnetometers.

\section{\label{sec:theory}Population and Coherence Magnetometry}

\begin{figure}[b]
\includegraphics[width=\linewidth]{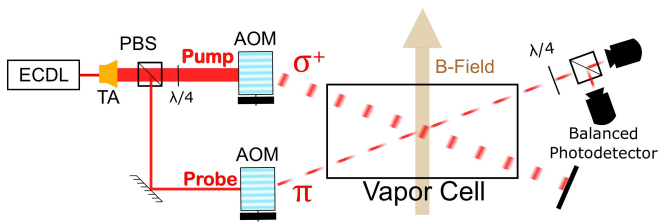}
\caption{\label{fig:popmag} (Color online) Schematic of time--domain population magnetometer.}
\end{figure}

Fig. \ref{fig:popmag} shows a schematic of a well--understood, time--domain,``population'' magnetometer \cite{Mlynek}. Here, a continuous wave (CW) diode laser is amplified by a tapered waveguide amplifier (TA) and used to generate a strong pump and a weak probe that are aligned at a small angle through a vapor cell containing an alkali sample such as rubidium. These beams are amplitude--modulated by acousto--optic modulators (AOMs). In this example, the circularly--polarized pump laser is tuned to the $^{85}$Rb $F=3\rightarrow F^\prime=4$ transition, and is used to optically pump atoms into the $F=3$, $m_F=3$ ground state magnetic sublevel, resulting in spin--polarization. If a magnetic field is applied perpendicular to the quantization axis defined by the pump laser, the transfer of population across the ground state manifold and back is modulated at the Larmor frequency $\omega_L = (\vec{\mu_F}\cdot\vec{B})/\hbar$. This population evolution is detected as a periodic variance in the differential transmission of the two orthogonal circular components of the probe beam.

The decay time of the signal---which is modulated at the Larmor frequency---is limited first by the transit time of the atoms through the pumping volume. If the pumping volume is extended to encompass the entire vapor cell, then the measurement time will be limited by the effect of wall collisions that decohere the Larmor oscillations. Although the measurement time can be extended using wall coatings, commonly--available coatings degrade at the high temperatures required for SERF magnetometry \cite{wallcoatings}. A simpler way to extend the measurement time is to add a high concentration of a \textit{buffer} gas---such as N$_2$ or a noble gas---whose principle requirement is a low spin--destruction cross--section. Collisions between the alkali atoms and the buffer gas will result in diffusive motion and effectively increase the transit time. Under these conditions the measurement time is limited by radiation trapping which scrambles the atomic polarization. The addition of a small concentration of \textit{quenching} gas---such as $\textrm{N}_2$---with a broad range of resonant energies can ensure that collisional de--excitations dominate spontaneous emission while preserving the spin--polarization. In this regime, spin--exchange collisions between rubidium atoms, which result in a transfer to atomic states that precess with the opposite phase, limit the time scale. Even so, this effect can be avoided by increasing the alkali density until the collisional frequency is large enough to re--initialize the phase of the Larmor precession resulting in the so--called SERF regime. 

Other transit time limited experiments involving the configuration in Fig.~\ref{fig:popmag} have been utilized for precise measurements of atomic $g$--factor ratios \cite{Iain2011,Kimball}. However, this type  of magnetometer is not ideally suited for diffusion measurements since the signal decay must be modelled by a complex function which is sensitive to various mechanisms of spin--depolarization in addition to diffusion. Further, the magnetic field response, which is also sensitive to various systematic effects, cannot be decoupled from the signal decay. In general, the transit time of atoms---$\tau_{transit}=f R^2/D$, where $R$ is the beam radius, $D$ is the diffusion coefficient, and $f$ is a form factor---is sensitive to the volume of the pump--probe overlap region. In devices of this type, the geometry of the overlapping region and its corresponding form factor contributes substantial errors into the diffusion measurement. The Rb--N$_2$ diffusion coefficient has been inferred by experiments such as this by illuminating an entire vapor cell, of simple geometry, with a circularly--polarized lamp source and monitoring the transmission \cite{Happer}. These measurements rely on knowledge of the form factor of the cell geometry and still must deconvolve magnetic field effects and all sources of spin de--polarization. 

\begin{figure}[b]
\includegraphics[width=\linewidth]{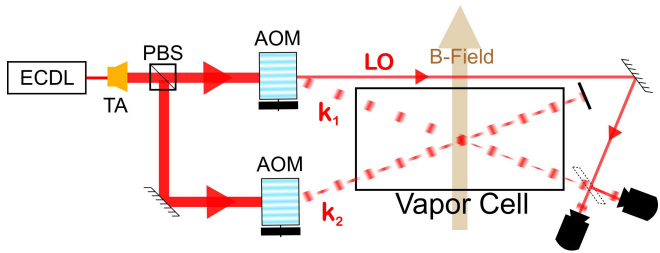}
\caption{\label{fig:cohmag} (Color online) Schematic of the coherence magnetometer.}
\end{figure}

\begin{figure}[b]
\includegraphics[width=\linewidth]{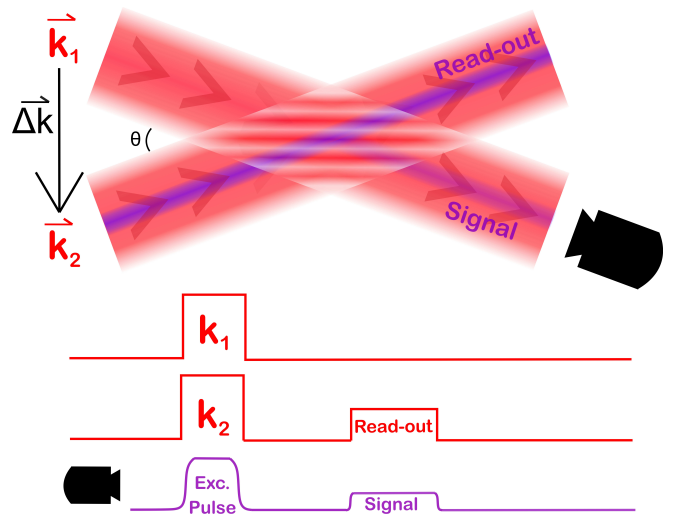}
\caption{\label{fig:grate} (Color online) Upper figure shows directions of the excitation pulses, read--out pulse, and signal. The lower figure shows the relative timing of the pulse envelopes along $\vec{k_1}$ and $\vec{k_2}$ and the signal envelopes recorded on the detector.}
\end{figure}

Fig. \ref{fig:cohmag} shows a schematic of a ``coherence" magnetometer. Here, a spatially--modulated coherence grating is created between adjacent magnetic sublevels of the $F=3$ ground state in $^{85}$Rb by an excitation pulse that consists of two perpendicular linear--polarized traveling waves, with wave vectors $\vec{k_1}$ and $\vec{k_2}$, aligned at a small angle $\theta$  (a few mrad). The grating is formed along the direction $\vec{\Delta k}=\vec{k_1}-\vec{k_2}$ as shown in Fig.~\ref{fig:grate} and has a spatial periodicity of $\sim\lambda/\theta$, where $\lambda=2\pi/k$ and $k$ is the magnitude of the wavevector $k=|\vec{k_1}|=|\vec{k_2}|$. The grating can be detected by applying a read--out pulse along the direction $\vec{k_2}$, and observing the coherent emission scattered along the phase--matched direction $\vec{k_1}$. This signal, called the magnetic grating free induction decay \mbox{(MGFID)} \cite{BermanLaserPhys}, exhibits a Gaussian decay with a time constant $\tau=2/ku\theta$, where $u$ is the most probable speed associated with the Maxwell--Boltzmann velocity distribution. This decay corresponds to the thermal motion of atoms causing the grating to dephase. The scattered electric field from the grating is then given by 

\begin{equation}
E(t)=E_0 e^{-(\frac{ku\theta}{2})^2t^2}  .
\label{eq:gauss}
\end{equation}{}

If the excitation pulses have opposite circular polarizations, they will excite coherences between magnetic sublevels separated by $\Delta m=2$. The signal scattered from this coherence grating will have the same time constant as for the case of perpendicular linear polarizations. The dephasing time of the grating has been used to measure the velocity distributions of warm vapor \cite{NYUVapor,Iain2008}, cold atomic gases \cite{NYUTrap,Iain2008}, and atomic beams \cite{Tonyushkin}.

In the presence of a magnetic field, the functional form of the coherence can have a complicated dependence, parameterized by the Larmor frequency. This behavior has been described in references \cite{NYUTrap,Iain2008} based on the formalism presented in reference \cite{Shore}. While Eq.~\ref{eq:gauss} assumes the thermal trajectory of the atoms is uninterrupted over the length scale of the grating, in the presence of a high concentration of buffer gas the mean--free--path of Rb atoms is reduced by collisions and may become much less than the grating spacing. In this limit, the motion of Rb atoms becomes a random walk that can be modelled by the diffusion equation \cite{BermanPRA94,ShimUnpublished,ShimThesis}. This condition is represented by 

\begin{equation}
\frac{\delta u}{\Gamma_{Col}}\ll\frac{1}{k\theta}.
\label{eq:diffregime}
\end{equation}

Here, $\delta u$ is the average velocity change per collision and $\Gamma_{Col}$ is the effective collisional rate. When Eq.~\ref{eq:diffregime} is satisfied, the evolution of the ground state density matrix $\rho$ can be described by the diffusion equation,

\begin{equation}
\frac{\partial\rho (x,t)}{\partial t}=-D\nabla^2\rho(x,t).
\label{eq:diffusion}
\end{equation}

Here, $D$ is the diffusion coefficient, which is inversely proportional to the perturber pressure. $D$ can be accurately converted to its value at atmospheric pressure $D_0$, using the relationship $D_0P_0=DP$. Here, $P_0$ is atmospheric pressure and $P$ is the buffer gas pressure in the experiment \cite{Chapman,cussler2009}. If the $x$--axis is along $\vec{\Delta k}$, the spatial dependence of the coherence $\rho$ may be written as $e^{ik\theta x}$. This results in

\begin{equation}
\frac{\partial\rho (x,t)}{\partial t}=-(\theta k)^2D\rho(x,t).
\label{eq:diffudiffer}
\end{equation}

The solution to Eq.~\ref{eq:diffudiffer} is a decaying exponential with a time constant $(k\theta)^2D$. The MGFID is therefore given by

\begin{equation}
E(t)=E_0 e^{-(\theta k)^2Dt}.
\label{eq:diffsig}
\end{equation}

Under these conditions, the coherent scattering from the grating is preserved but the signal exhibits an exponential decay with a characteristic time constant $\tau=1/(D(k\theta)^2)$. Since $(k\theta)^{-1}$ represents the characteristic length scale in this problem, namely the grating spacing, the scaling law for $\tau$ is representative of a random walk. Therefore, the coherence magnetometer offers a direct approach for measuring diffusion rates \cite{ShimUnpublished}. However, this method is prone to inaccuracies since the scattered signal has a small amplitude and is sensitive to magnetic field gradients.

As a result, we have exploited an interesting aspect of the lin--perp--lin excitation, namely that it simultaneously produces a density grating with the same period as the coherence grating. Accordingly, we are able to record decays with much improved signal--to--noise ratios and with greater accuracy due to the insensitivity of the density grating to magnetic fields and field gradients. It should be noted that the density gratings used in this work can be modelled without atomic recoil or matter--wave interference effects \cite{CAHNPrl,websiteAMOreview,Atoms,brynlePRA,carsonPRA}. By recording the decay time as a function of angle, we rely on Eq.~\ref{eq:diffsig} to measure the diffusion coefficient with a statistical uncertainty of 1\%. The novelty of the technique and the high precision are the central results of this paper.

The density grating forms as the result of a spatially--periodic light intensity modulation in the combined (standing--wave) excitation field \cite{channeling}. This standing--wave potential channels atoms into the nodes of the optical potential since our excitation pulses, which are resonant with the unperturbed rubidium resonance, are effectively blue--detuned with respect to the center of the collisionally--broadened line. This is because collisions red--shift the center of the atomic resonance by $\sim$3 GHz for our experimental conditions \cite{RomalisBroad,CollBroad} due to the influence of the buffer gas potential \cite{Corney}. The channeling of atoms to the nodes produces a grating that has a larger contrast than the coherence grating. This is evident from a comparison of the signal strengths associated with both gratings. We believe that the coherence grating has a smaller contrast since there is no explicit optical pumping stage in this experiment. Instead, the population imbalances, required to generate a coherence, are caused by spontaneous emission and collisional redistribution during the excitation pulse.

\section{\label{sec:Experiment}Experimental Details}

\begin{figure*}
\includegraphics[width=\textwidth]{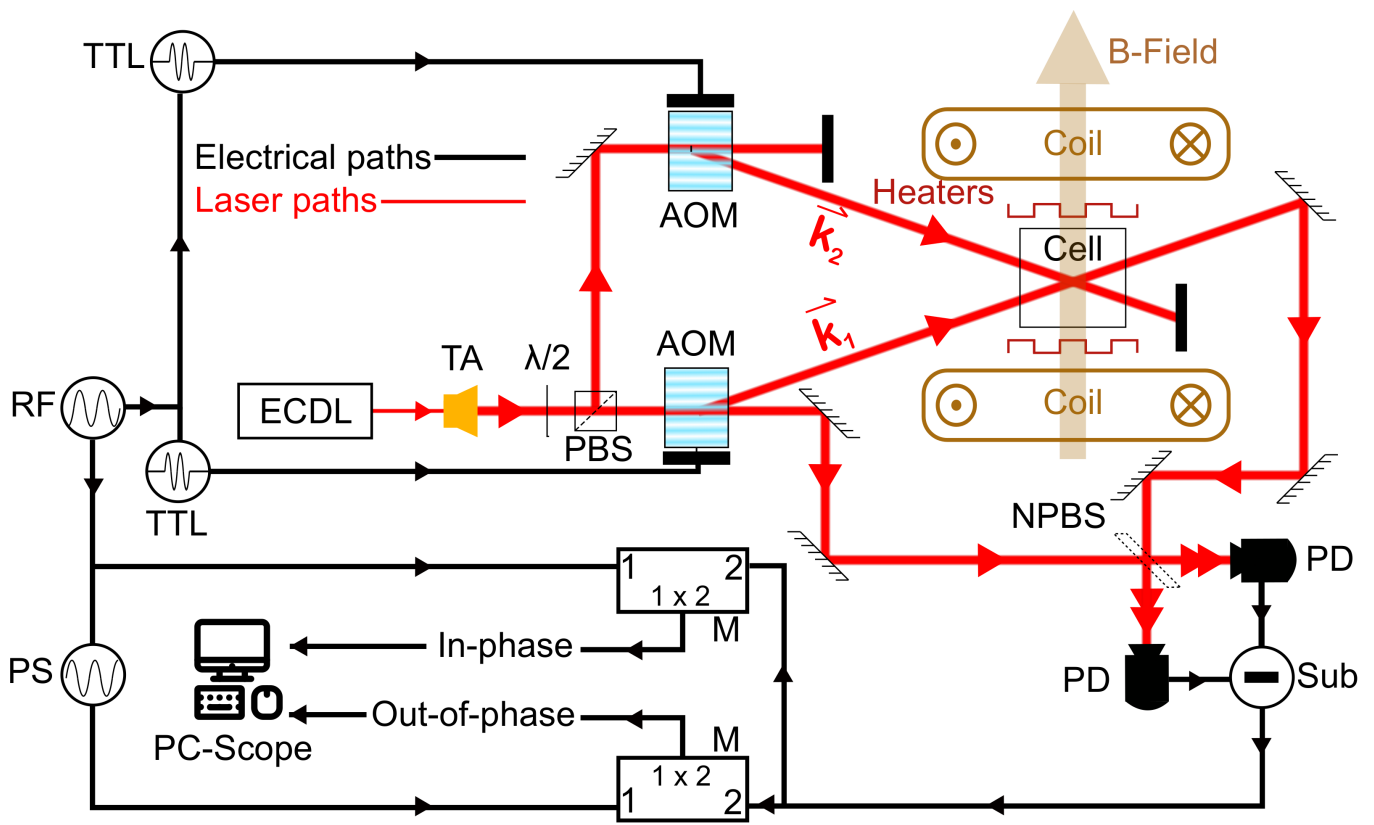}
\caption{\label{fig:fullsketch} (Color online) Schematic of the experimental set--up. The laser (ECDL) seeds the TA, the output of which is split by half wave plate ($\lambda /2$)---polarizing beam splitter (PBS) and sent to the two AOMs. The two AOMs are driven by $80~\textrm{MHz}$ pulses produced by the RF generator (RF), and their respective TTL switches. In the diffusion measurements, the coils minimize the external field ($B\approx0$). The beams are overlapped on a non-polarizing beam splitter (NPBS) forming the heterodyne signal recorded on the photodiodes (PD). These signals are subtracted to remove amplitude fluctuations. The subtracted signal is mixed down on mixers (M) to DC using the RF frequency, and the RF frequency with a 90$\degree$ phase shift provided by the phase--shifter (PS). The in--phase and out--of--phase components of the signal are obtained in this manner. These mixed--down signals are recorded and analyzed on the PC oscilloscope.}
\end{figure*}

The experiment relies on a home--built external--cavity diode laser (ECDL) \cite{HerminaRSI} that seeds a TA with $\sim15~\textrm{mW}$ of light to realize an output of 2 W \cite{SPIEPWest}. The ECDL is frequency stabilized with respect to the $F=2 \rightarrow F'=2,3$ crossover peak in $^{87}\textrm{Rb}$ using saturated absorption in a 5 cm--long vapor cell. The output of the TA is split into two beams, each amplitude--modulated by an $80~\textrm{MHz}$ AOM as shown in Fig.~\ref{fig:fullsketch}. The AOMs are driven by an RF network consisting of an RF generator, RF amplifiers, TTL switches and pulse generators. By adjusting the power and timing of the RF pulses, the power and pulse sequence of the diffracted AOM output may be varied. The upshifted beams from these AOMs, which are at a frequency of $53~\textrm{MHz}$ below the $F=2 \rightarrow F'=3$ resonance, are aligned along directions $\vec{k_1}$ and $\vec{k_2}$ (see Fig.~\ref{fig:grate}) through a $10~\textrm{cm}$, quartz, vapor cell containing isotopically pure $^{87}\textrm{Rb}$ and $564\pm2~\textrm{Torr}$ of $\textrm{N}_2$. The pressure in the sealed cell was spectroscopically determined as reported in Sec.~\ref{sec:pmeas}. The isotopic purity of the cell simplifies its magnetic response but is not required for the eventual diffusion measurement which is insensitive to magnetic fields. The cell is insulated and maintained at a temperature of $50\pm2\degree~C$ using a resistive heater. As shown in Fig.~\ref{fig:fullsketch}, the cell is placed in a constant magnetic field, transverse to the direction of laser propagation. The B--field is produced by a pair of ``racetrack" coils with an elliptical cross--section. The undiffracted beam from the $k_1$ AOM, at a frequency $80~\textrm{MHz}$ below the frequency of the diffracted beam, bypasses the cell and is combined with the beam along $\vec{k_1}$ on a beam splitter downstream from the cell. The outputs of the beamsplitter, that contain a heterodyne signal with a beat frequency of $80~\textrm{MHz}$, are incident on a balanced detector. This detector consists of two Si:PIN photodiodes with $1~\textrm{ns}$ risetimes that are biased to produce signals with opposite polarity. The combined, $80~\textrm{MHz}$ signal from the photodiodes is amplified and mixed down to DC to generate the in--phase and out--of--phase components. These signal envelopes are further amplified and recorded on a 12--bit analog to digital converter (ADC) with a bandwidth of $125~\textrm{MHz}$ corresponding to a two--channel acquisition rate of $250~\textrm{MS}/\textrm{s}$. The total amplitude is obtained by adding the two signal components in quadrature. The experiment is operated at approximately 1 kHz repetition rates using digital delay generators. The time base of these generators is slaved to a $10~\textrm{MHz}$ rubidium atomic clock \cite{stanfordResearch} with an Allan deviation floor value of $3 \times 10^{-13}$ at one hour. The pulses from the delay generators are coupled to the AOMs using TTL switches. The RF generator that produces the $80~\textrm{MHz}$ AOM drive frequency is also phase locked to the same $10~\textrm{MHz}$ output of the rubidium clock. This practice ensures phase noise makes a negligible contribution to the error of this measurement. 

The same setup was used to generate signals from the coherence grating (MGFID) and the density grating (as shown in Fig.~\ref{fig:cohmag}. The quartz vapor cell was replaced with a pyrex cell containing a natural abundance of $^{85}$Rb and $^{87}$Rb isotopes and no buffer gas to record reference MGFID signals. The important difference between this cell and the previously mentioned cell is the absence of a buffer gas. For studies of the population magnetometer, which were carried out in the quartz cell containing isotopically pure $^{87}$Rb vapor, the $k_2$ beam was circularly--polarized and served as the pump, while the $k_1$ beam was linearly polarized and attenuated to serve as the probe. The signal was recorded by measuring the differential absorption of the oppositely polarized circular components of the probe beam that are split by the $\lambda/4$ waveplate--cube beam splitter combination shown in Fig.~\ref{fig:popmag}. In all of these experiments the decay of the signal is recorded by varying the delay time of the read--out or probe pulses, and recording the average of 80--100 repetitions. 

The angle between $\vec{k_1}$ and $\vec{k_2}$ was measured using a scanning knife edge profiler with a rotation frequency of $\sim$10~Hz. The separation between the beams was measured at two locations separated by $\sim$2~m. The center of each beam was located using Gaussian fits to the profiler output, allowing the separation between beams to be determined. The variation in the rotation frequency as a function of time was characterized by an Allan deviation plot. The errors in the frequency and separation were then propagated and combined in quadrature to obtain the error in the angle. To ensure the condition for diffusion (Eq.~\ref{eq:diffregime}) was met for the 564~Torr cell, the selected angles ranged from 1.5 to 9~mrad, corresponding to grating spacings of 10 to 80~$\mu$m, more than two orders of magnitude greater than the mean--free--path for 564~Torr of N$_2$ ($\sim280$~nm).

\section{\label{sec:results}Results and Discussion}

\begin{figure*}
\includegraphics[width=\textwidth]{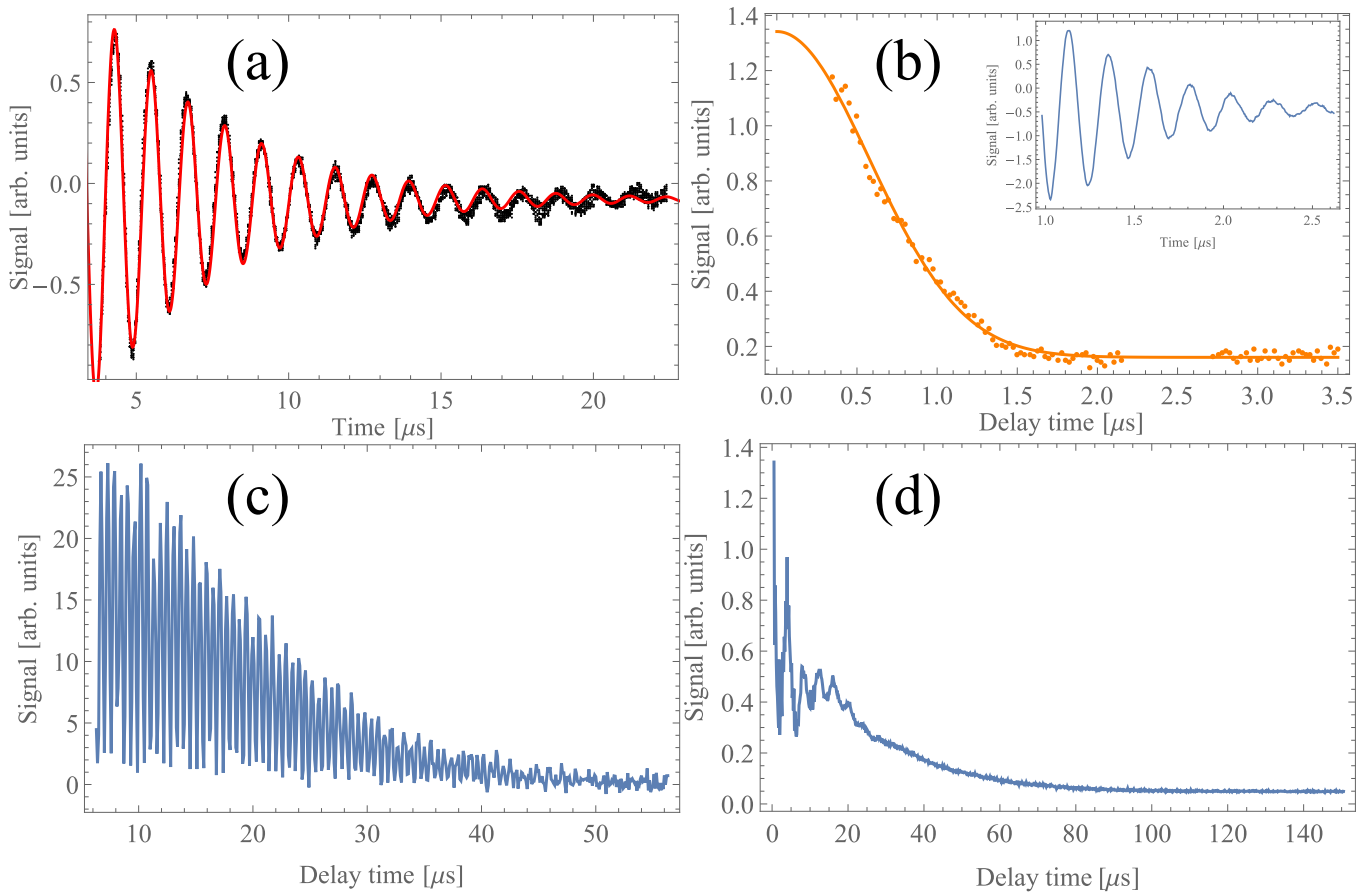}
\caption{\label{fig:signals} (Color online) Representative responses of the coherence and population magnetometers. Field strengths are varied to clearly demonstrate the magnetic field response over the varied timescales. (a) Decay of the spin--polarization of a population magnetometer in the pyrex vapour cell containing natural abundance of Rb and no buffer gas. With a magnetic field of $0.2 ~\textrm{G}$, the observed Larmor frequency ($0.433~ \textrm{MHz}$) is consistent with this field. (b) Gaussian decay of the total signal of the coherence magnetometer in the same cell. The magnetic field is zeroed to minimize Larmor oscillations in the decay and infer the most probable speed. The excitation pulse width is $70~\textrm{ns}$ and $\theta=2$ mrad. The fit gives $u=246~ \textrm{m}/\textrm{s}$, which corresponds to a temperature of $30\degree\textrm{C}$, agreeing with the cell temperature. Inset shows the MGFID signal in a magnetic field of $13.8 ~\textrm{G}$. The Larmor oscillation frequency is $6.5 ~\textrm{MHz}$. (c) The response of the coherence magnetometer in a magnetic field of $1.25~ \textrm{G}$ with 564~Torr of $\textrm{N}_2$ buffer gas. The excitation pulse width is 1 $\mu \textrm{s}$, which ensures that the signal from the population grating is small. The Larmor oscillation of $1.7~ \textrm{MHz}$ is consistent with the applied magnetic field. (d) The response of the coherence magnetometer with a longer excitation pulse ($80 ~\mu \textrm{s}$), $\theta = 5.23$ mrad, and magnetic field of $0.17 ~\textrm{G}$. The magnetic field response is visible at early times but the coherent scattering from the density grating dominates at later times.}
\end{figure*}

Fig. \ref{fig:signals} shows representative signals of the coherent transients described in this paper. Fig. \ref{fig:signals}(a) shows the population magnetometer signal recorded in the pyrex vapor cell without a buffer gas. The duration of the pump pulse was $300~\textrm{ns}$ and the duration of the weak probe pulse was $100~\mu \textrm{s}$. Here, the signal represents the intensity of the differential absorption of the two polarization components of the probe pulse. The decay time is principally limited by the estimated transit time across the $3~\textrm{mm}\times3~\textrm{mm}$ probe beam (approximately $12~\mu \textrm{s}$). The frequency of the Larmor oscillations is consistent with the applied magnetic field. The data illustrates the difficulty of using this signal for measurements of diffusion. Firstly, the magnetic field response must be deconvolved. Secondly, the decay time is sensitive to the probe volume, which must be quantified. These requirements add significant uncertainty to any diffusion measurement. Fig. \ref{fig:signals}(b) shows the MGFID of the coherence magnetometer recorded in the same cell as in \ref{fig:signals}(a), with $k_1$ and $k_2$ excitation pulse widths of $70~\textrm{ns}$. Each point was recorded by varying the delay of an intense, $70~\textrm{ns}$ read--out pulse, then integrating and adding the two components in quadrature to obtain the total intensity. These measurements were carried out by annulling the ambient magnetic field to avoid Larmor oscillations from the coherence grating. A Gaussian fit to the intensity gives a temperature of $30\degree\textrm{C}$ which is consistent with the cell temperature. The $1/\theta$ dependence of the decay time has been verified in reference \cite{ShimUnpublished}. The inset shows the in--phase component of the MGFID signal in a magnetic field ($13.8~\textrm{G}$). The Larmor oscillation frequency of $6.5 ~\textrm{MHz}$ is consistent with the expected value for this field. The inset data was taken with a single, long, weak read--out pulse, as was done for Fig.~\ref{fig:signals}(a).

Fig. \ref{fig:signals}(c) shows the MGFID signal in the presence of a magnetic field of $1.25~\textrm{G}$, from isotopically pure $^{87}$Rb vapor in 564 Torr of N$_2$ gas. Here, the $k_1$ and $k_2$ excitation pulse widths were 1 $\mu \textrm{s}$. The delay time of a $100~\textrm{ns}$, intense, read--out pulse, was varied to record the signal decay. As in Fig.~\ref{fig:signals}(b), the total intensity of the scattered signal is displayed. The frequency of the Larmor oscillations ($1.7~\textrm{MHz})$ is once again consistent with the applied magnetic field. Although this signal could be used for a diffusion measurement, it is necessary to remove the effects of the magnetic field and residual field gradients to obtain smooth decays. This process can be error prone, and can only be avoided by good magnetic shielding. However, a density grating is insensitive to magnetic field effects and does not have shielding requirements. Additionally, as we will show, a much larger signal--to--noise ratio can be realized with a density grating. In Fig.~\ref{fig:signals}(d) the width of the excitation pulse is extended to 80~$\mu \textrm{s}$ and the magnetic field is reduced to $0.17 ~\textrm{G}$. The coherent scattering is recorded in the same manner as in Figs. \ref{fig:signals}(b) and \ref{fig:signals}(c) by varying the delay time of an intense $100~\textrm{ns}$ read--out pulse. The coherent scattering from the magnetic field--dependent coherence grating is visible at early time delays while the scattering from the density grating dominates at later times. It is evident that the signal from the density grating can be observed on much longer timescales since it has a greater amplitude than the signal from the coherence grating for suitably long excitation pulses. 

\begin{figure}
\includegraphics[width=\linewidth]{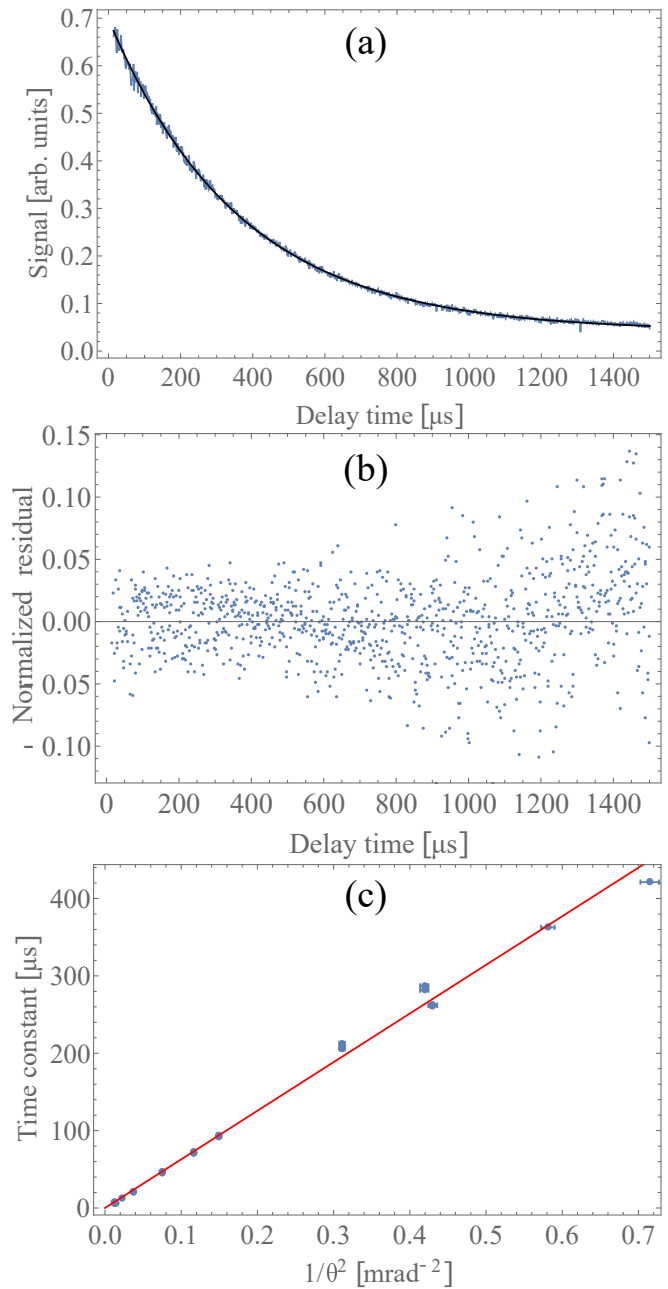}
\caption{\label{fig:diffusion} (Color online) (a) Exponential signal decay of a density grating at 564 Torr $\textrm{N}_2$. The excitation pulse width is $80 ~\mu \textrm{s}$ and the delay of a $100~\textrm{ns}$ read--out pulse is varied. Here, each excitation pulse has a single--photon, average Rabi frequency $\Omega\approx10\Gamma$, where $\Gamma$ is the radiative rate of the excited state. The dataset consists of 1000 points. Here, $\theta=1.31\pm0.01$ mrad and the decay time $\tau = 363\pm1 ~\mu \textrm{s}$, which together give $D=0.246~\textrm{cm}^2/\textrm{s}$. (b) Normalized fit residuals of the exponential decay. (c) Fit to the decay time of the density grating as a function of $1/\theta^2$ showing a linear dependence. The slope gives $D$ = $0.245\pm0.002~\textrm{cm}/\textrm{s}^2$, which represents a statistical error of 1\%. Scaling to atmospheric pressure gives $D_0 = 0.1819\pm0.0024~\textrm{cm}^2/\textrm{s}$.}
\end{figure}

Fig. \ref{fig:diffusion}(a) shows the signal from a long--lived density grating recorded by varying the delay time of an intense $100~\textrm{ns}$ read--out pulse. This data was recorded over a timescale of $\sim$1 hour by randomizing 1000 delay times. Over this duration we verified that the uncertainty in $\theta$ is less than $0.001$~mrad by sampling the beam profile on a profile analyzer. The signal exhibits the expected exponential decay curve represented by the fit line. Here, the angle $\theta$ between $\vec{k_1}$ and $\vec{k_2}$ was adjusted to be $1.31\pm0.01$ mrad. The magnetic fields were also minimized using a single pair of coils to reduce the amplitude of Larmor oscillations that are visible at the beginning of the curve. These kinds of decay curves were used in the diffusion measurement. The measured decay time constant $\tau$ is $363~\mu \textrm{s}$ with a fit error of $\pm1~\mu \textrm{s}$. The data also shows that the timescale of the decay is comparable to those obtained in all previous diffusion measurements. The value of $D$ (at 564~Torr) extracted from the fit on the basis of Eq.~\ref{eq:diffsig} is $0.247\pm0.008~ \textrm{cm}^2/\textrm{s}$. Fig. \ref{fig:diffusion}(b) shows the normalized fit residuals which demonstrate that the model based on an exponential fit agrees with the data. 

Fig. \ref{fig:diffusion}(c) shows the decay time constant measured as a function of angle $\theta$ and plotted as a function of $1/\theta^2$. This trend demonstrates one of the key advantages of this technique, namely the ability to change the length scale on which diffusion occurs. We note that this length scale (the grating spacing) is significantly smaller than the beam diameter of 3 mm. Over the range of angles this ratio of the beam size to grating spacing varies from 27 to 360, suggesting that the transit time correction is negligible. The linear dependence confirms the characteristic scaling law expected for a diffusion--dominated system (Eq.~\ref{eq:diffsig}). The slope of this line gives $D=0.245\pm0.002~\textrm{cm}^2/\textrm{s}$, which represents a statistical uncertainty of 1\%. We scale this to atmospheric pressure and obtain $D_0= 0.1819~\textrm{cm}^2/\textrm{s}$. The combined error in $k$, $\theta$ and $\tau$, computed in quadrature, is 0.9\%, which consistent with the observed statistical error. Additionally, the variation in cell temperature is 0.5\%. These factors contribute to an overall uncertainty of 1.3\% (assuming a $T^{3/2}$ scaling law for D \cite{Chapman,cussler2009}), giving an absolute error of $\pm 0.0024~\textrm{cm}^2/\textrm{s}$. We now address the most challenging source of systematic uncertainty, associated with the pressure in the cell.

\subsection{Pressure Measurement}
\label{sec:pmeas}
\begin{figure}
\includegraphics[width=0.9\linewidth]{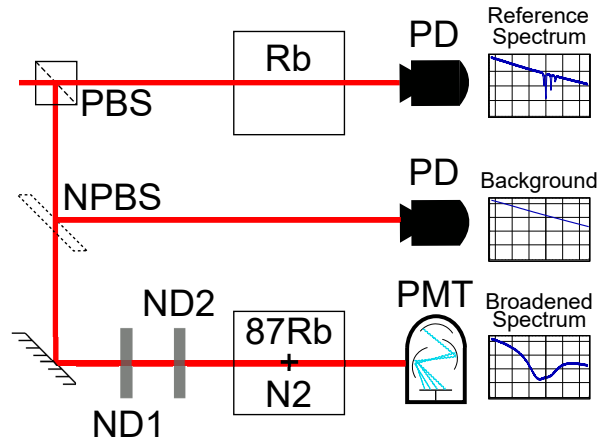}
\caption{\label{fig:presdiagram} (Color online) Schematic of experimental set up to measure the pressure in the isotopically purified $^{87}$Rb cell. A diode laser is scanned over $\sim60$~GHz by scanning the current. The power of the laser is attenuated using two 13~dB ND filters to avoid spectral distortions due to optical pumping. The frequency scan is calibrated using the spectrum of a low--pressure reference cell. Power fluctuations in the laser output are calibrated by recording the laser intensity on a photodiode. The pressure--broadened spectrum of the $^{87}$Rb cell is recorded using a PMT.}
\end{figure}

The cell pressure provided by the manufacturer was measured with a capacitance manometer with a precision of $\pm5\%$ but the uncertainty in the pressure at the time of heating and separation from the gas manifold is estimated to be around $\pm10\%$. To reduce the systematic uncertainty in the pressure specification, we carried out an independent spectroscopic measurement of pressure using the setup shown in Fig.~\ref{fig:presdiagram}. Recent examples of pressure broadening studies in alkali vapors relevant to magnetometry are described in references~\cite{CollBroad,RomalisBroad,Broad2013}. In the experiment, we scanned a free--running laser diode with an estimated linewidth of 50 MHz across the pressure--broadened (and pressure--shifted) resonances in the isotopically purified $^{87}$Rb cell. The cell was maintained at a temperature of 47$\degree$C as measured by an array of thermocouples. These parameters are similar to the conditions at which the diffusion measurement was recorded (50$\degree$C). At this temperature, the increase in rubidium density compared to room temperature, as measured by the area under the absorption spectra, was consistent with estimates from rubidium vapor pressure curves~\cite{Corney,RbVP}. 

\begin{figure}
\includegraphics[width=0.9\linewidth]{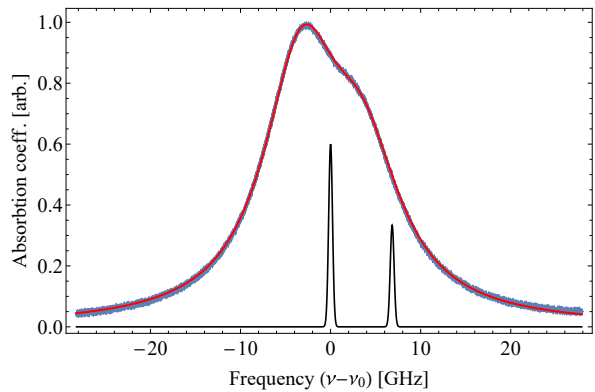}
\caption{\label{fig:doublehump} (Color online) Absorption spectrum of the $^{87}$Rb cell near the D2 transition, $\nu_0$ denotes the vacuum resonance from the $F=2$ ground state. Background subtracted data is shown in light blue with the fit shown in red, noise in data has been emphasized to distinguish the data from the fit. The pressure inferred from this fit is $561.0 \pm 0.43$~Torr, corresponding to a Lorentzian width of $10.267 \pm0.008$~GHz, and a shift of $-3.25\pm0.003$~GHz. A representation of the low--pressure, Doppler--broadened spectrum associated with the D2 transitions from the $F=1$ and $F=2$ ground states in $^{87}$Rb (solid black) is also shown to demonstrate the broadening and shift caused by collisions (the amplitude change is arbitrary) }
\end{figure}

The spectra from the isotopically purified $^{87}$Rb cell were recorded on a photmultiplier tube (PMT) for a range of logarithmically stepped laser powers from $1-2~\mu\textrm{W}$ at which there are no detectable optical pumping effects. The laser diode was scanned at rates of $21-106~\textrm{Hz}$ with $10,000-30,000$ samples per sweep. The spectrum was obtained by averaging 560-1660 sweeps and a representative sample is shown in Fig.~\ref{fig:doublehump}. To provide a frequency calibration, a reference scan was simultaneously recorded in a vapour cell containing a natural isotopic abundance of Rb at room temperature. A photodiode was used to record the laser intensity variation during the scan. This information was used to model frequency dependant intensity variations of the laser diode, such as those caused by etalon effects and the diode gain curve. 

To infer the N$_2$ gas concentration from the measured spectrum, we perform a fit to the collisionally broadened and shifted profile of $^{87}$Rb. The profile is modeled as the sum of two Voigt profiles \cite{Corney} separated by the known hyperfine splitting of the $^{87}$Rb ground states (6.823468 GHz)~\cite{Steck87}. Both Voigt profiles share the same Lorentzian width and shift parameters, which are defined by a single pressure parameter that uses the relationships measured in reference~\cite{CollBroad}. The ratio of peak heights of the two Voigt profiles are assumed to be fixed. We also use a scalable background term based on the measured laser intensity. To ensure that the fit parameters are strongly constrained, datasets obtained by increasing and decreasing the laser frequency during the scan are fit simultaneously. The pressure is extracted from fits such as the one shown in Fig.~\ref{fig:doublehump}. This fit is superimposed on an illustration showing the Doppler-broadened $^{87}$Rb spectrum from a cell without buffer gas.

\begin{figure}
\includegraphics[width=0.9\linewidth]{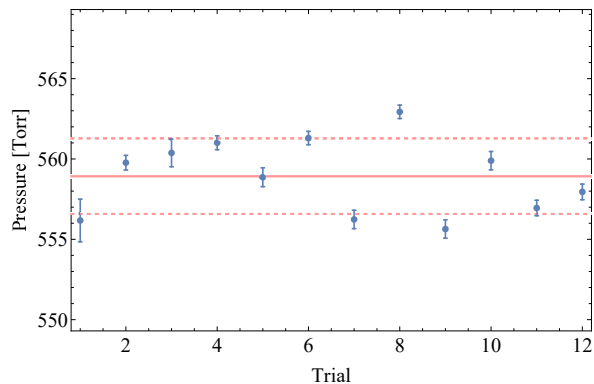}
\caption{\label{fig:Pvals} (Color online) Distribution of pressure measurements taken at varying input laser powers (between 1-2 $\mu W$) and laser scan rates of 106~Hz or 21~Hz, at an effective temperature of 47$\degree$C. The fit errors in the pressure parameter of the Voigt profiles are shown as error bars. The additional spread is attributed to the ~$\pm0.5\%$ temperature fluctuations in the cell. The mean value of 559.0~Torr is indicated by a pink line. The standard deviation of $\pm2.3$~Torr is indicated by the dashed pink lines.}
\end{figure}

Fig.~\ref{fig:Pvals} shows the inferred pressure obtained with laser powers ranging from $1-2~ \mu\textrm{W}$ and an effective temperature of 47$\degree$C. We find that the fit errors typically range from 0.07$\%$ to 0.2$\%$. The scatter in the data can be attributed to temperature fluctuations of $\pm 1.5 \degree$C over the time in which the data was acquired. The average value was determined to be $559\pm 2~\textrm{Torr}$ at 47$\degree$C. Using the ideal gas law, this value can be scaled to the cell temperature during the diffusion measurement (50$\degree$C) giving a pressure of 564$\pm$2~Torr. This is lower than the bound provided by the cell manufacturer when scaled to a temperature of 50$\degree$C ($650 \pm 65$~Torr). It is significant that the statistical uncertainty in the pressure measurement has reduced the dominant systematic uncertainty in the measurement of the diffusion coefficient from 10\% to 0.4\%, resulting in a negligible contribution to the overall error.

The final value, $D_0=0.1819\pm0.0024~\textrm{cm}^2/s$, reported in Table~\ref{tab:diff} has been rescaled using $D_0=D'\frac{P'}{P_0}$, where $D'$ is the diffusion coefficient measured at 50$\degree$C and $P'$ is the pressure inside the cell at this temperature, inferred from the pressure measurement at 47$\degree$C. Here, $D_0$ and $P_0$ represent the values at atmospheric pressure.

\subsection*{Conclusions}

We have demonstrated a distinctive and accurate measurement of the diffusion coefficient of Rb in N$_2$ relevant to magnetometry. Ideally, the systematic effect due to the N$_2$ concentration should be measured at buffer gas concentrations of several atmospheres so that spectra can be fit to a smooth lineshape as in reference~\cite{RomalisBroad,Broad2013}. Since the buffer gas pressure in our isotopically purified rubidium cell could not be changed, we fit the pressure--broadened spectrum to a function appropriate for our pressure range where the hyperfine splitting is significant. From Table \ref{tab:diff} it is clear that our measurement disagrees with the previous most precise measurement obtained using spin echoes \cite{Ishikawa}. Although the diffusion coefficients measured by modelling optical pumping curves \cite{Franz,Happer,Wagshul,EricksonThesis} are in good agreement with our results, we note that these measurements do not report error bounds.

The disagreement between our measurement and reference \cite{Ishikawa} could point to unaccounted systematic effects in either technique. We also note that our inferred diffusion coefficients depend on the low--pressure measurements of the broadening and shift parameters in reference~\cite{CollBroad}, which are in disagreement with the same parameters determined at much higher pressure~\cite{RomalisBroad,Broad2013}. However, we note that using the values in references~\cite{RomalisBroad,Broad2013} would also result in a discrepancy with respect to the diffusion coefficient measurement in reference~\cite{Ishikawa}. 

Systematic errors in this work that have not been explicitly accounted for include the residual effects of wall collisions at the cell windows and the temperature scaling law used to rescale $D_0$ to compare with other works in Table~\ref{tab:diff}, which depends on the nature of the intermolecular potential \cite{NIST,Chapman}. This measurement can be improved further by reducing the uncertainty in the angle measurement using a more stable spatial profiler, and by increasing the number of measurements to improve statistics. It should also be possible to carry out this experiment in a gas manifold in which the pressure of the buffer gas can be varied and measured precisely using both spectroscopy and a capacitance manometer. This technique can also be extended to other buffer gases and alkali vapors used in magnetometry for further comparisons with theoretical models.

\section{\label{sec:acks}Acknowledgements}
We acknowledge helpful discussions with Louis Marmet. We also acknowledge helpful discussions with Paul Berman of the University of Michigan in the early stages of this work. This  work was  supported  by  the  Canada  Foundation  for Innovation,  the Ontario  Innovation  Trust,  the  Ontario  Centers  of Excellence, the Natural Sciences and Engineering Research Council of Canada, and York University.


\begin{thebibliography}{54}%
\makeatletter
\providecommand \@ifxundefined [1]{%
 \@ifx{#1\undefined}
}%
\providecommand \@ifnum [1]{%
 \ifnum #1\expandafter \@firstoftwo
 \else \expandafter \@secondoftwo
 \fi
}%
\providecommand \@ifx [1]{%
 \ifx #1\expandafter \@firstoftwo
 \else \expandafter \@secondoftwo
 \fi
}%
\providecommand \natexlab [1]{#1}%
\providecommand \enquote  [1]{``#1''}%
\providecommand \bibnamefont  [1]{#1}%
\providecommand \bibfnamefont [1]{#1}%
\providecommand \citenamefont [1]{#1}%
\providecommand \href@noop [0]{\@secondoftwo}%
\providecommand \href [0]{\begingroup \@sanitize@url \@href}%
\providecommand \@href[1]{\@@startlink{#1}\@@href}%
\providecommand \@@href[1]{\endgroup#1\@@endlink}%
\providecommand \@sanitize@url [0]{\catcode `\\12\catcode `\$12\catcode
  `\&12\catcode `\#12\catcode `\^12\catcode `\_12\catcode `\%12\relax}%
\providecommand \@@startlink[1]{}%
\providecommand \@@endlink[0]{}%
\providecommand \url  [0]{\begingroup\@sanitize@url \@url }%
\providecommand \@url [1]{\endgroup\@href {#1}{\urlprefix }}%
\providecommand \urlprefix  [0]{URL }%
\providecommand \Eprint [0]{\href }%
\providecommand \doibase [0]{https://doi.org/}%
\providecommand \selectlanguage [0]{\@gobble}%
\providecommand \bibinfo  [0]{\@secondoftwo}%
\providecommand \bibfield  [0]{\@secondoftwo}%
\providecommand \translation [1]{[#1]}%
\providecommand \BibitemOpen [0]{}%
\providecommand \bibitemStop [0]{}%
\providecommand \bibitemNoStop [0]{.\EOS\space}%
\providecommand \EOS [0]{\spacefactor3000\relax}%
\providecommand \BibitemShut  [1]{\csname bibitem#1\endcsname}%
\let\auto@bib@innerbib\@empty
\bibitem [{\citenamefont {Allred}\ \emph {et~al.}(2002)\citenamefont {Allred},
  \citenamefont {Lyman}, \citenamefont {Kornack},\ and\ \citenamefont
  {Romalis}}]{Allred}%
  \BibitemOpen
  \bibfield  {author} {\bibinfo {author} {\bibfnamefont {J.~C.}\ \bibnamefont
  {Allred}}, \bibinfo {author} {\bibfnamefont {R.~N.}\ \bibnamefont {Lyman}},
  \bibinfo {author} {\bibfnamefont {T.~W.}\ \bibnamefont {Kornack}},\ and\
  \bibinfo {author} {\bibfnamefont {M.~V.}\ \bibnamefont {Romalis}},\
  }\bibfield  {title} {\bibinfo {title} {High-sensitivity atomic magnetometer
  unaffected by spin-exchange relaxation},\ }\href
  {https://doi.org/10.1103/PhysRevLett.89.130801} {\bibfield  {journal}
  {\bibinfo  {journal} {Phys. Rev. Lett.}\ }\textbf {\bibinfo {volume} {89}},\
  \bibinfo {pages} {130801} (\bibinfo {year} {2002})}\BibitemShut {NoStop}%
\bibitem [{\citenamefont {Fagaly}(2006)}]{SQUID}%
  \BibitemOpen
  \bibfield  {author} {\bibinfo {author} {\bibfnamefont {R.~L.}\ \bibnamefont
  {Fagaly}},\ }\bibfield  {title} {\bibinfo {title} {Superconducting quantum
  interference device instruments and applications},\ }\href
  {https://doi.org/10.1063/1.2354545} {\bibfield  {journal} {\bibinfo
  {journal} {Rev. Sci. Inst.}\ }\textbf {\bibinfo {volume} {77}},\ \bibinfo
  {pages} {101101} (\bibinfo {year} {2006})}\BibitemShut {NoStop}%
\bibitem [{\citenamefont {Budker}\ and\ \citenamefont
  {Romalis}(2007)}]{SERFReview}%
  \BibitemOpen
  \bibfield  {author} {\bibinfo {author} {\bibfnamefont {D.}~\bibnamefont
  {Budker}}\ and\ \bibinfo {author} {\bibfnamefont {M.}~\bibnamefont
  {Romalis}},\ }\bibfield  {title} {\bibinfo {title} {Optical magnetometry},\
  }\href {https://doi.org/10.1038/nphys566} {\bibfield  {journal} {\bibinfo
  {journal} {Nature Physics}\ }\textbf {\bibinfo {volume} {3}},\ \bibinfo
  {pages} {227} (\bibinfo {year} {2007})}\BibitemShut {NoStop}%
\bibitem [{\citenamefont {Suter}\ \emph {et~al.}(1990)\citenamefont {Suter},
  \citenamefont {Rosatzin},\ and\ \citenamefont {Mlynek}}]{Mlynek}%
  \BibitemOpen
  \bibfield  {author} {\bibinfo {author} {\bibfnamefont {D.}~\bibnamefont
  {Suter}}, \bibinfo {author} {\bibfnamefont {M.}~\bibnamefont {Rosatzin}},\
  and\ \bibinfo {author} {\bibfnamefont {J.}~\bibnamefont {Mlynek}},\
  }\bibfield  {title} {\bibinfo {title} {Optically driven spin nutations in the
  ground state of atomic sodium},\ }\href
  {https://doi.org/10.1103/PhysRevA.41.1634} {\bibfield  {journal} {\bibinfo
  {journal} {Phys. Rev. A}\ }\textbf {\bibinfo {volume} {41}},\ \bibinfo
  {pages} {1634} (\bibinfo {year} {1990})}\BibitemShut {NoStop}%
\bibitem [{\citenamefont {Berman}\ \emph {et~al.}(1993)\citenamefont {Berman},
  \citenamefont {Rogers},\ and\ \citenamefont
  {Dubetsky}}]{bermanDensityMatrix}%
  \BibitemOpen
  \bibfield  {author} {\bibinfo {author} {\bibfnamefont {P.~R.}\ \bibnamefont
  {Berman}}, \bibinfo {author} {\bibfnamefont {G.}~\bibnamefont {Rogers}},\
  and\ \bibinfo {author} {\bibfnamefont {B.}~\bibnamefont {Dubetsky}},\
  }\bibfield  {title} {\bibinfo {title} {Rate equations between
  electronic-state manifolds},\ }\href
  {https://doi.org/10.1103/PhysRevA.48.1506} {\bibfield  {journal} {\bibinfo
  {journal} {Phys. Rev. A}\ }\textbf {\bibinfo {volume} {48}},\ \bibinfo
  {pages} {1506} (\bibinfo {year} {1993})}\BibitemShut {NoStop}%
\bibitem [{\citenamefont {Barrett}(2012)}]{brynle}%
  \BibitemOpen
  \bibfield  {author} {\bibinfo {author} {\bibfnamefont {B.}~\bibnamefont
  {Barrett}},\ }\emph {\bibinfo {title} {Techniques for Measuring the Atomic
  Recoil Frequency using a Grating-Echo Atom Interferometer}},\ \href@noop {}
  {\bibinfo {type} {{PhD} dissertation}},\ \bibinfo  {school} {York
  University}, \bibinfo {address} {Department of Physics and Astronomy}
  (\bibinfo {year} {2012})\BibitemShut {NoStop}%
\bibitem [{\citenamefont {Pouliot}\ \emph
  {et~al.}(2018{\natexlab{a}})\citenamefont {Pouliot}, \citenamefont {Beica},
  \citenamefont {Carew}, \citenamefont {Vorozcovs}, \citenamefont {Carlse},
  \citenamefont {Barrett},\ and\ \citenamefont {Kumarakrishnan}}]{spiedefense}%
  \BibitemOpen
  \bibfield  {author} {\bibinfo {author} {\bibfnamefont {A.}~\bibnamefont
  {Pouliot}}, \bibinfo {author} {\bibfnamefont {H.~C.}\ \bibnamefont {Beica}},
  \bibinfo {author} {\bibfnamefont {A.}~\bibnamefont {Carew}}, \bibinfo
  {author} {\bibfnamefont {A.}~\bibnamefont {Vorozcovs}}, \bibinfo {author}
  {\bibfnamefont {G.}~\bibnamefont {Carlse}}, \bibinfo {author} {\bibfnamefont
  {B.}~\bibnamefont {Barrett}},\ and\ \bibinfo {author} {\bibfnamefont
  {A.}~\bibnamefont {Kumarakrishnan}},\ }\bibfield  {title} {\bibinfo {title}
  {{Investigations of optical pumping for magnetometry using an auto-locking
  laser system}},\ }in\ \href {https://doi.org/10.1117/12.2304598} {\emph
  {\bibinfo {booktitle} {Laser Technology for Defense and Security XIV}}},\
  Vol.\ \bibinfo {volume} {10637},\ \bibinfo {editor} {edited by\ \bibinfo
  {editor} {\bibfnamefont {M.}~\bibnamefont {Dubinskiy}}\ and\ \bibinfo
  {editor} {\bibfnamefont {T.~C.}\ \bibnamefont {Newell}}},\ \bibinfo
  {organization} {International Society for Optics and Photonics}\ (\bibinfo
  {publisher} {SPIE},\ \bibinfo {year} {2018})\ pp.\ \bibinfo {pages} {40 --
  47}\BibitemShut {NoStop}%
\bibitem [{\citenamefont {Happer}(1972)}]{HapperOptPump}%
  \BibitemOpen
  \bibfield  {author} {\bibinfo {author} {\bibfnamefont {W.}~\bibnamefont
  {Happer}},\ }\bibfield  {title} {\bibinfo {title} {Optical pumping},\ }\href
  {https://doi.org/10.1103/RevModPhys.44.169} {\bibfield  {journal} {\bibinfo
  {journal} {Rev. Mod. Phys.}\ }\textbf {\bibinfo {volume} {44}},\ \bibinfo
  {pages} {169} (\bibinfo {year} {1972})}\BibitemShut {NoStop}%
\bibitem [{\citenamefont {Ben-Amar~Baranga}\ \emph {et~al.}(1998)\citenamefont
  {Ben-Amar~Baranga}, \citenamefont {Appelt}, \citenamefont {Romalis},
  \citenamefont {Erickson}, \citenamefont {Young}, \citenamefont {Cates},\ and\
  \citenamefont {Happer}}]{Baranga}%
  \BibitemOpen
  \bibfield  {author} {\bibinfo {author} {\bibfnamefont {A.}~\bibnamefont
  {Ben-Amar~Baranga}}, \bibinfo {author} {\bibfnamefont {S.}~\bibnamefont
  {Appelt}}, \bibinfo {author} {\bibfnamefont {M.~V.}\ \bibnamefont {Romalis}},
  \bibinfo {author} {\bibfnamefont {C.~J.}\ \bibnamefont {Erickson}}, \bibinfo
  {author} {\bibfnamefont {A.~R.}\ \bibnamefont {Young}}, \bibinfo {author}
  {\bibfnamefont {G.~D.}\ \bibnamefont {Cates}},\ and\ \bibinfo {author}
  {\bibfnamefont {W.}~\bibnamefont {Happer}},\ }\bibfield  {title} {\bibinfo
  {title} {Polarization of ${}^{3}\mathrm{He}$ by spin exchange with optically
  pumped $\mathrm{Rb}$ and $\mathrm{K}$ vapors},\ }\href
  {https://doi.org/10.1103/PhysRevLett.80.2801} {\bibfield  {journal} {\bibinfo
   {journal} {Phys. Rev. Lett.}\ }\textbf {\bibinfo {volume} {80}},\ \bibinfo
  {pages} {2801} (\bibinfo {year} {1998})}\BibitemShut {NoStop}%
\bibitem [{\citenamefont {Jenkins}(1968)}]{Jenkins}%
  \BibitemOpen
  \bibfield  {author} {\bibinfo {author} {\bibfnamefont {D.~R.}\ \bibnamefont
  {Jenkins}},\ }\bibfield  {title} {\bibinfo {title} {The determination of
  cross sections for the quenching of resonance radiation of metal atoms.
  $\mathrm{II}$. results for potassium, rubidium and caesium},\ }\href
  {http://www.jstor.org/stable/2415911} {\bibfield  {journal} {\bibinfo
  {journal} {Proc. R. Soc. London, Ser. A}\ }\textbf {\bibinfo {volume}
  {303}},\ \bibinfo {pages} {453} (\bibinfo {year} {1968})}\BibitemShut
  {NoStop}%
\bibitem [{\citenamefont {Speller}\ \emph {et~al.}(1979)\citenamefont
  {Speller}, \citenamefont {Staudenmayer},\ and\ \citenamefont
  {Kempter}}]{Speller}%
  \BibitemOpen
  \bibfield  {author} {\bibinfo {author} {\bibfnamefont {E.}~\bibnamefont
  {Speller}}, \bibinfo {author} {\bibfnamefont {B.}~\bibnamefont
  {Staudenmayer}},\ and\ \bibinfo {author} {\bibfnamefont {V.}~\bibnamefont
  {Kempter}},\ }\bibfield  {title} {\bibinfo {title} {Quenching cross sections
  for alkali-inert gas collisions},\ }\href
  {https://doi.org/10.1007/BF01408379} {\bibfield  {journal} {\bibinfo
  {journal} {Z. Phys. A}\ }\textbf {\bibinfo {volume} {291}},\ \bibinfo {pages}
  {311} (\bibinfo {year} {1979})}\BibitemShut {NoStop}%
\bibitem [{\citenamefont {Wagshul}\ and\ \citenamefont
  {Chupp}(1994)}]{Wagshul}%
  \BibitemOpen
  \bibfield  {author} {\bibinfo {author} {\bibfnamefont {M.~E.}\ \bibnamefont
  {Wagshul}}\ and\ \bibinfo {author} {\bibfnamefont {T.~E.}\ \bibnamefont
  {Chupp}},\ }\bibfield  {title} {\bibinfo {title} {Laser optical pumping of
  high-density $\mathrm{Rb}$ in polarized $^{3}\mathrm{He}$ targets},\ }\href
  {https://doi.org/10.1103/PhysRevA.49.3854} {\bibfield  {journal} {\bibinfo
  {journal} {Phys. Rev. A}\ }\textbf {\bibinfo {volume} {49}},\ \bibinfo
  {pages} {3854} (\bibinfo {year} {1994})}\BibitemShut {NoStop}%
\bibitem [{\citenamefont {Allard}\ and\ \citenamefont
  {Kielkopf}(1982)}]{allardkielkopf}%
  \BibitemOpen
  \bibfield  {author} {\bibinfo {author} {\bibfnamefont {N.}~\bibnamefont
  {Allard}}\ and\ \bibinfo {author} {\bibfnamefont {J.}~\bibnamefont
  {Kielkopf}},\ }\bibfield  {title} {\bibinfo {title} {The effect of neutral
  nonresonant collisions on atomic spectral lines},\ }\href
  {https://doi.org/10.1103/RevModPhys.54.1103} {\bibfield  {journal} {\bibinfo
  {journal} {Rev. Mod. Phys.}\ }\textbf {\bibinfo {volume} {54}},\ \bibinfo
  {pages} {1103} (\bibinfo {year} {1982})}\BibitemShut {NoStop}%
\bibitem [{\citenamefont {Zeng}\ \emph {et~al.}(1985)\citenamefont {Zeng},
  \citenamefont {Wu}, \citenamefont {Call}, \citenamefont {Miron},
  \citenamefont {Schreiber},\ and\ \citenamefont {Happer}}]{Happer}%
  \BibitemOpen
  \bibfield  {author} {\bibinfo {author} {\bibfnamefont {X.}~\bibnamefont
  {Zeng}}, \bibinfo {author} {\bibfnamefont {Z.}~\bibnamefont {Wu}}, \bibinfo
  {author} {\bibfnamefont {T.}~\bibnamefont {Call}}, \bibinfo {author}
  {\bibfnamefont {E.}~\bibnamefont {Miron}}, \bibinfo {author} {\bibfnamefont
  {D.}~\bibnamefont {Schreiber}},\ and\ \bibinfo {author} {\bibfnamefont
  {W.}~\bibnamefont {Happer}},\ }\bibfield  {title} {\bibinfo {title}
  {Experimental determination of the rate constants for spin exchange between
  optically pumped $\mathrm{K}$, $\mathrm{Rb}$, and $\mathrm{Cs}$ atoms and
  $^{129}\mathrm{Xe}$ nuclei in alkali-metal--noble-gas van der
  $\mathrm{W}$aals molecules},\ }\href
  {https://doi.org/10.1103/PhysRevA.31.260} {\bibfield  {journal} {\bibinfo
  {journal} {Phys. Rev. A}\ }\textbf {\bibinfo {volume} {31}},\ \bibinfo
  {pages} {260} (\bibinfo {year} {1985})}\BibitemShut {NoStop}%
\bibitem [{\citenamefont {Ishikawa}\ and\ \citenamefont
  {Yabuzaki}(2000)}]{Ishikawa}%
  \BibitemOpen
  \bibfield  {author} {\bibinfo {author} {\bibfnamefont {K.}~\bibnamefont
  {Ishikawa}}\ and\ \bibinfo {author} {\bibfnamefont {T.}~\bibnamefont
  {Yabuzaki}},\ }\bibfield  {title} {\bibinfo {title} {Diffusion coefficient
  and sublevel coherence of $\mathrm{Rb}$ atoms in $\mathrm{N}_{2}$ buffer
  gas},\ }\href {https://doi.org/10.1103/PhysRevA.62.065401} {\bibfield
  {journal} {\bibinfo  {journal} {Phys. Rev. A}\ }\textbf {\bibinfo {volume}
  {62}},\ \bibinfo {pages} {065401} (\bibinfo {year} {2000})}\BibitemShut
  {NoStop}%
\bibitem [{\citenamefont {Walker}\ and\ \citenamefont
  {Happer}(1997)}]{HapperPolImg}%
  \BibitemOpen
  \bibfield  {author} {\bibinfo {author} {\bibfnamefont {T.~G.}\ \bibnamefont
  {Walker}}\ and\ \bibinfo {author} {\bibfnamefont {W.}~\bibnamefont
  {Happer}},\ }\bibfield  {title} {\bibinfo {title} {Spin-exchange optical
  pumping of noble-gas nuclei},\ }\href
  {https://doi.org/10.1103/RevModPhys.69.629} {\bibfield  {journal} {\bibinfo
  {journal} {Rev. Mod. Phys.}\ }\textbf {\bibinfo {volume} {69}},\ \bibinfo
  {pages} {629} (\bibinfo {year} {1997})}\BibitemShut {NoStop}%
\bibitem [{\citenamefont {Higbie}\ \emph {et~al.}(2011)\citenamefont {Higbie},
  \citenamefont {Rochester}, \citenamefont {Patton}, \citenamefont
  {Holzlöhner}, \citenamefont {Bonaccini~Calia},\ and\ \citenamefont
  {Budker}}]{budkerSodium}%
  \BibitemOpen
  \bibfield  {author} {\bibinfo {author} {\bibfnamefont {J.~M.}\ \bibnamefont
  {Higbie}}, \bibinfo {author} {\bibfnamefont {S.~M.}\ \bibnamefont
  {Rochester}}, \bibinfo {author} {\bibfnamefont {B.}~\bibnamefont {Patton}},
  \bibinfo {author} {\bibfnamefont {R.}~\bibnamefont {Holzlöhner}}, \bibinfo
  {author} {\bibfnamefont {D.}~\bibnamefont {Bonaccini~Calia}},\ and\ \bibinfo
  {author} {\bibfnamefont {D.}~\bibnamefont {Budker}},\ }\bibfield  {title}
  {\bibinfo {title} {Magnetometry with mesospheric sodium},\ }\href@noop {}
  {\bibfield  {journal} {\bibinfo  {journal} {Proc. Natl. Acad. Sci. USA}\
  }\textbf {\bibinfo {volume} {108}},\ \bibinfo {pages} {3522–3525} (\bibinfo
  {year} {2011})}\BibitemShut {NoStop}%
\bibitem [{\citenamefont {McNeal}(1962)}]{McNeal}%
  \BibitemOpen
  \bibfield  {author} {\bibinfo {author} {\bibfnamefont {R.~J.}\ \bibnamefont
  {McNeal}},\ }\bibfield  {title} {\bibinfo {title} {Disorientation cross
  sections in optical pumping},\ }\href {https://doi.org/10.1063/1.1733089}
  {\bibfield  {journal} {\bibinfo  {journal} {J. Chem. Phys.}\ }\textbf
  {\bibinfo {volume} {37}},\ \bibinfo {pages} {2726} (\bibinfo {year}
  {1962})}\BibitemShut {NoStop}%
\bibitem [{\citenamefont {Franz}\ and\ \citenamefont
  {Sooriamoorthi}(1973)}]{Franz}%
  \BibitemOpen
  \bibfield  {author} {\bibinfo {author} {\bibfnamefont {F.~A.}\ \bibnamefont
  {Franz}}\ and\ \bibinfo {author} {\bibfnamefont {C.~E.}\ \bibnamefont
  {Sooriamoorthi}},\ }\bibfield  {title} {\bibinfo {title} {Analytic
  expressions for transient signals in the optical pumping of alkali-metal
  vapors},\ }\href {https://doi.org/10.1103/PhysRevA.8.2390} {\bibfield
  {journal} {\bibinfo  {journal} {Phys. Rev. A}\ }\textbf {\bibinfo {volume}
  {8}},\ \bibinfo {pages} {2390} (\bibinfo {year} {1973})}\BibitemShut
  {NoStop}%
\bibitem [{\citenamefont {Erickson}(2000)}]{EricksonThesis}%
  \BibitemOpen
  \bibfield  {author} {\bibinfo {author} {\bibfnamefont {C.~J.}\ \bibnamefont
  {Erickson}},\ }\href@noop {} {\bibinfo {type} {{PhD} dissertation}},\
  \bibinfo  {school} {Princeton University} (\bibinfo {year}
  {2000})\BibitemShut {NoStop}%
\bibitem [{\citenamefont {Parniak}\ and\ \citenamefont
  {Wasilewski}(2014)}]{Parniak}%
  \BibitemOpen
  \bibfield  {author} {\bibinfo {author} {\bibfnamefont {M.}~\bibnamefont
  {Parniak}}\ and\ \bibinfo {author} {\bibfnamefont {W.}~\bibnamefont
  {Wasilewski}},\ }\bibfield  {title} {\bibinfo {title} {Direct observation of
  atomic diffusion in warm rubidium ensembles},\ }\href
  {https://doi.org/10.1007/s00340-013-5712-y} {\bibfield  {journal} {\bibinfo
  {journal} {Appl. Phys. B}\ }\textbf {\bibinfo {volume} {116}},\ \bibinfo
  {pages} {415} (\bibinfo {year} {2014})}\BibitemShut {NoStop}%
\bibitem [{\citenamefont {Ishikawa}(2016)}]{Ishikawa2}%
  \BibitemOpen
  \bibfield  {author} {\bibinfo {author} {\bibfnamefont {K.}~\bibnamefont
  {Ishikawa}},\ }\bibfield  {title} {\bibinfo {title} {Spin-polarized lithium
  diffusion in a glass hot-vapor cell},\ }\href
  {https://doi.org/10.1007/s00340-016-6505-x} {\bibfield  {journal} {\bibinfo
  {journal} {Applied Physics B}\ }\textbf {\bibinfo {volume} {122}},\ \bibinfo
  {pages} {224} (\bibinfo {year} {2016})}\BibitemShut {NoStop}%
\bibitem [{\citenamefont {Aoki}\ and\ \citenamefont {Mitsui}(2016)}]{Aoki}%
  \BibitemOpen
  \bibfield  {author} {\bibinfo {author} {\bibfnamefont {K.}~\bibnamefont
  {Aoki}}\ and\ \bibinfo {author} {\bibfnamefont {T.}~\bibnamefont {Mitsui}},\
  }\bibfield  {title} {\bibinfo {title} {Observing random walks of atoms in
  buffer gas through resonant light absorption},\ }\href
  {https://doi.org/10.1103/PhysRevA.94.012703} {\bibfield  {journal} {\bibinfo
  {journal} {Phys. Rev. A}\ }\textbf {\bibinfo {volume} {94}},\ \bibinfo
  {pages} {012703} (\bibinfo {year} {2016})}\BibitemShut {NoStop}%
\bibitem [{\citenamefont {Rothberg}\ and\ \citenamefont
  {Bloembergen}(1984)}]{bloembergen}%
  \BibitemOpen
  \bibfield  {author} {\bibinfo {author} {\bibfnamefont {L.~J.}\ \bibnamefont
  {Rothberg}}\ and\ \bibinfo {author} {\bibfnamefont {N.}~\bibnamefont
  {Bloembergen}},\ }\bibfield  {title} {\bibinfo {title} {High-resolution
  four-wave light-mixing studies of collision-induced coherence in na vapor},\
  }\href {https://doi.org/10.1103/PhysRevA.30.820} {\bibfield  {journal}
  {\bibinfo  {journal} {Phys. Rev. A}\ }\textbf {\bibinfo {volume} {30}},\
  \bibinfo {pages} {820} (\bibinfo {year} {1984})}\BibitemShut {NoStop}%
\bibitem [{\citenamefont {Chapman}\ \emph {et~al.}(1990)\citenamefont
  {Chapman}, \citenamefont {Cowling},\ and\ \citenamefont {D.}}]{Chapman}%
  \BibitemOpen
  \bibfield  {author} {\bibinfo {author} {\bibfnamefont {S.}~\bibnamefont
  {Chapman}}, \bibinfo {author} {\bibfnamefont {T.}~\bibnamefont {Cowling}},\
  and\ \bibinfo {author} {\bibfnamefont {B.}~\bibnamefont {D.}},\ }\href@noop
  {} {\emph {\bibinfo {title} {The Mathematical Theory of Non-uniform Gases: An
  Account of the Kinetic Theory of Viscosity, Thermal Conduction and Diffusion
  in Gases}}},\ \bibinfo {edition} {3rd}\ ed.\ (\bibinfo  {publisher}
  {Cambridge University Press},\ \bibinfo {year} {1990})\BibitemShut {NoStop}%
\bibitem [{\citenamefont {Cussler}(2009)}]{cussler2009}%
  \BibitemOpen
  \bibfield  {author} {\bibinfo {author} {\bibfnamefont {E.~L.}\ \bibnamefont
  {Cussler}},\ }\bibinfo {title} {Contents},\ in\ \href@noop {} {\emph
  {\bibinfo {booktitle} {Diffusion: Mass Transfer in Fluid Systems}}},\
  \bibinfo {series and number} {Cambridge Series in Chemical Engineering}\
  (\bibinfo  {publisher} {Cambridge University Press},\ \bibinfo {year}
  {2009})\ pp.\ \bibinfo {pages} {vii--xii},\ \bibinfo {edition} {3rd}\
  ed.\BibitemShut {Stop}%
\bibitem [{\citenamefont {Seltzer}\ and\ \citenamefont
  {Romalis}(2009)}]{wallcoatings}%
  \BibitemOpen
  \bibfield  {author} {\bibinfo {author} {\bibfnamefont {S.~J.}\ \bibnamefont
  {Seltzer}}\ and\ \bibinfo {author} {\bibfnamefont {M.~V.}\ \bibnamefont
  {Romalis}},\ }\bibfield  {title} {\bibinfo {title} {High-temperature alkali
  vapor cells with antirelaxation surface coatings},\ }\href
  {https://doi.org/10.1063/1.3236649} {\bibfield  {journal} {\bibinfo
  {journal} {J. Appl. Phys.}\ }\textbf {\bibinfo {volume} {106}},\ \bibinfo
  {pages} {114905} (\bibinfo {year} {2009})}\BibitemShut {NoStop}%
\bibitem [{\citenamefont {Chan}\ \emph {et~al.}(2011)\citenamefont {Chan},
  \citenamefont {Barrett},\ and\ \citenamefont {Kumarakrishnan}}]{Iain2011}%
  \BibitemOpen
  \bibfield  {author} {\bibinfo {author} {\bibfnamefont {I.}~\bibnamefont
  {Chan}}, \bibinfo {author} {\bibfnamefont {B.}~\bibnamefont {Barrett}},\ and\
  \bibinfo {author} {\bibfnamefont {A.}~\bibnamefont {Kumarakrishnan}},\
  }\bibfield  {title} {\bibinfo {title} {Precise determination of atomic
  $g$-factor ratios from a dual isotope magneto-optical trap},\ }\href
  {https://doi.org/10.1103/PhysRevA.84.032509} {\bibfield  {journal} {\bibinfo
  {journal} {Phys. Rev. A}\ }\textbf {\bibinfo {volume} {84}},\ \bibinfo
  {pages} {032509} (\bibinfo {year} {2011})}\BibitemShut {NoStop}%
\bibitem [{\citenamefont {Mora}\ \emph {et~al.}(2019)\citenamefont {Mora},
  \citenamefont {Cobos}, \citenamefont {Fuentes},\ and\ \citenamefont
  {Jackson~Kimball}}]{Kimball}%
  \BibitemOpen
  \bibfield  {author} {\bibinfo {author} {\bibfnamefont {J.}~\bibnamefont
  {Mora}}, \bibinfo {author} {\bibfnamefont {A.}~\bibnamefont {Cobos}},
  \bibinfo {author} {\bibfnamefont {D.}~\bibnamefont {Fuentes}},\ and\ \bibinfo
  {author} {\bibfnamefont {D.~F.}\ \bibnamefont {Jackson~Kimball}},\ }\bibfield
   {title} {\bibinfo {title} {Measurement of the ratio between g-factors of the
  ground states of $^{87}\mathrm{Rb}$ and $^{85}\mathrm{Rb}$},\ }\href
  {https://doi.org/10.1002/andp.201800281} {\bibfield  {journal} {\bibinfo
  {journal} {Ann. Phys. (Leipzig)}\ }\textbf {\bibinfo {volume} {531}},\
  \bibinfo {pages} {1800281} (\bibinfo {year} {2019})}\BibitemShut {NoStop}%
\bibitem [{\citenamefont {Berman}\ and\ \citenamefont
  {Dubetsky}(1994)}]{BermanLaserPhys}%
  \BibitemOpen
  \bibfield  {author} {\bibinfo {author} {\bibfnamefont {P.~R.}\ \bibnamefont
  {Berman}}\ and\ \bibinfo {author} {\bibfnamefont {B.}~\bibnamefont
  {Dubetsky}},\ }\bibfield  {title} {\bibinfo {title} {Magnetic grating free
  induction decay and magnetic grating echo},\ }\href@noop {} {\bibfield
  {journal} {\bibinfo  {journal} {Laser Physics}\ }\textbf {\bibinfo {volume}
  {4}},\ \bibinfo {pages} {1017} (\bibinfo {year} {1994})}\BibitemShut
  {NoStop}%
\bibitem [{\citenamefont {Kumarakrishnan}\ \emph
  {et~al.}(1998{\natexlab{a}})\citenamefont {Kumarakrishnan}, \citenamefont
  {Shim}, \citenamefont {Cahn},\ and\ \citenamefont {Sleator}}]{NYUVapor}%
  \BibitemOpen
  \bibfield  {author} {\bibinfo {author} {\bibfnamefont {A.}~\bibnamefont
  {Kumarakrishnan}}, \bibinfo {author} {\bibfnamefont {U.}~\bibnamefont
  {Shim}}, \bibinfo {author} {\bibfnamefont {S.~B.}\ \bibnamefont {Cahn}},\
  and\ \bibinfo {author} {\bibfnamefont {T.}~\bibnamefont {Sleator}},\
  }\bibfield  {title} {\bibinfo {title} {Ground-state grating echoes from
  $\mathrm{Rb}$ vapor at room temperature},\ }\href
  {https://doi.org/10.1103/PhysRevA.58.3868} {\bibfield  {journal} {\bibinfo
  {journal} {Phys. Rev. A}\ }\textbf {\bibinfo {volume} {58}},\ \bibinfo
  {pages} {3868} (\bibinfo {year} {1998}{\natexlab{a}})}\BibitemShut {NoStop}%
\bibitem [{\citenamefont {Chan}\ \emph {et~al.}(2008)\citenamefont {Chan},
  \citenamefont {Andreyuk}, \citenamefont {Beattie}, \citenamefont {Barrett},
  \citenamefont {Mok}, \citenamefont {Weel},\ and\ \citenamefont
  {Kumarakrishnan}}]{Iain2008}%
  \BibitemOpen
  \bibfield  {author} {\bibinfo {author} {\bibfnamefont {I.}~\bibnamefont
  {Chan}}, \bibinfo {author} {\bibfnamefont {A.}~\bibnamefont {Andreyuk}},
  \bibinfo {author} {\bibfnamefont {S.}~\bibnamefont {Beattie}}, \bibinfo
  {author} {\bibfnamefont {B.}~\bibnamefont {Barrett}}, \bibinfo {author}
  {\bibfnamefont {C.}~\bibnamefont {Mok}}, \bibinfo {author} {\bibfnamefont
  {M.}~\bibnamefont {Weel}},\ and\ \bibinfo {author} {\bibfnamefont
  {A.}~\bibnamefont {Kumarakrishnan}},\ }\bibfield  {title} {\bibinfo {title}
  {Properties of magnetic sublevel coherences for precision measurements},\
  }\href {https://doi.org/10.1103/PhysRevA.78.033418} {\bibfield  {journal}
  {\bibinfo  {journal} {Phys. Rev. A}\ }\textbf {\bibinfo {volume} {78}},\
  \bibinfo {pages} {033418} (\bibinfo {year} {2008})}\BibitemShut {NoStop}%
\bibitem [{\citenamefont {Kumarakrishnan}\ \emph
  {et~al.}(1998{\natexlab{b}})\citenamefont {Kumarakrishnan}, \citenamefont
  {Cahn}, \citenamefont {Shim},\ and\ \citenamefont {Sleator}}]{NYUTrap}%
  \BibitemOpen
  \bibfield  {author} {\bibinfo {author} {\bibfnamefont {A.}~\bibnamefont
  {Kumarakrishnan}}, \bibinfo {author} {\bibfnamefont {S.~B.}\ \bibnamefont
  {Cahn}}, \bibinfo {author} {\bibfnamefont {U.}~\bibnamefont {Shim}},\ and\
  \bibinfo {author} {\bibfnamefont {T.}~\bibnamefont {Sleator}},\ }\bibfield
  {title} {\bibinfo {title} {Magnetic grating echoes from laser-cooled atoms},\
  }\href {https://doi.org/10.1103/PhysRevA.58.R3387} {\bibfield  {journal}
  {\bibinfo  {journal} {Phys. Rev. A}\ }\textbf {\bibinfo {volume} {58}},\
  \bibinfo {pages} {R3387} (\bibinfo {year} {1998}{\natexlab{b}})}\BibitemShut
  {NoStop}%
\bibitem [{\citenamefont {Tonyushkin}\ \emph {et~al.}(2010)\citenamefont
  {Tonyushkin}, \citenamefont {Kumarakrishnan}, \citenamefont {Turlapov},\ and\
  \citenamefont {Sleator}}]{Tonyushkin}%
  \BibitemOpen
  \bibfield  {author} {\bibinfo {author} {\bibfnamefont {A.}~\bibnamefont
  {Tonyushkin}}, \bibinfo {author} {\bibfnamefont {A.}~\bibnamefont
  {Kumarakrishnan}}, \bibinfo {author} {\bibfnamefont {A.}~\bibnamefont
  {Turlapov}},\ and\ \bibinfo {author} {\bibfnamefont {T.}~\bibnamefont
  {Sleator}},\ }\bibfield  {title} {\bibinfo {title} {Magnetic coherence
  gratings in a high-flux atomic beam},\ }\href
  {https://doi.org/10.1140/epjd/e2010-00085-8} {\bibfield  {journal} {\bibinfo
  {journal} {The European Physical Journal D}\ }\textbf {\bibinfo {volume}
  {58}},\ \bibinfo {pages} {39} (\bibinfo {year} {2010})}\BibitemShut {NoStop}%
\bibitem [{\citenamefont {Shore}(1990)}]{Shore}%
  \BibitemOpen
  \bibfield  {author} {\bibinfo {author} {\bibfnamefont {B.}~\bibnamefont
  {Shore}},\ }\href {https://books.google.ca/books?id=0ldXAQAACAAJ} {\emph
  {\bibinfo {title} {The Theory of Coherent Atomic Excitation, 2 Volume
  Set}}},\ The Theory of Coherent Atomic Excitation\ (\bibinfo  {publisher}
  {Wiley},\ \bibinfo {year} {1990})\BibitemShut {NoStop}%
\bibitem [{\citenamefont {Berman}(1994)}]{BermanPRA94}%
  \BibitemOpen
  \bibfield  {author} {\bibinfo {author} {\bibfnamefont {P.~R.}\ \bibnamefont
  {Berman}},\ }\bibfield  {title} {\bibinfo {title} {Collisional decay and
  revival of the grating stimulated echo},\ }\href
  {https://doi.org/10.1103/PhysRevA.49.2922} {\bibfield  {journal} {\bibinfo
  {journal} {Phys. Rev. A}\ }\textbf {\bibinfo {volume} {49}},\ \bibinfo
  {pages} {2922} (\bibinfo {year} {1994})}\BibitemShut {NoStop}%
\bibitem [{\citenamefont {Shim}\ \emph {et~al.}(2005)\citenamefont {Shim},
  \citenamefont {Kumarakrishnan}, \citenamefont {Turlapov}, \citenamefont
  {Cahn},\ and\ \citenamefont {Sleator}}]{ShimUnpublished}%
  \BibitemOpen
  \bibfield  {author} {\bibinfo {author} {\bibfnamefont {U.}~\bibnamefont
  {Shim}}, \bibinfo {author} {\bibfnamefont {A.}~\bibnamefont
  {Kumarakrishnan}}, \bibinfo {author} {\bibfnamefont {A.~V.}\ \bibnamefont
  {Turlapov}}, \bibinfo {author} {\bibfnamefont {S.~B.}\ \bibnamefont {Cahn}},\
  and\ \bibinfo {author} {\bibfnamefont {T.}~\bibnamefont {Sleator}},\
  }\bibfield  {title} {\bibinfo {title} {Collisional revival of magnetic
  grating free induction decay}} (\bibinfo {year} {2005}),\ \bibinfo {note}
  {(Unpublished)}\BibitemShut {NoStop}%
\bibitem [{\citenamefont {Shim}(1997)}]{ShimThesis}%
  \BibitemOpen
  \bibfield  {author} {\bibinfo {author} {\bibfnamefont {U.}~\bibnamefont
  {Shim}},\ }\href@noop {} {\bibinfo {type} {{PhD} dissertation}},\ \bibinfo
  {school} {New York University} (\bibinfo {year} {1997})\BibitemShut {NoStop}%
\bibitem [{\citenamefont {Cahn}\ \emph {et~al.}(1997)\citenamefont {Cahn},
  \citenamefont {Kumarakrishnan}, \citenamefont {Shim}, \citenamefont
  {Sleator}, \citenamefont {Berman},\ and\ \citenamefont {Dubetsky}}]{CAHNPrl}%
  \BibitemOpen
  \bibfield  {author} {\bibinfo {author} {\bibfnamefont {S.~B.}\ \bibnamefont
  {Cahn}}, \bibinfo {author} {\bibfnamefont {A.}~\bibnamefont
  {Kumarakrishnan}}, \bibinfo {author} {\bibfnamefont {U.}~\bibnamefont
  {Shim}}, \bibinfo {author} {\bibfnamefont {T.}~\bibnamefont {Sleator}},
  \bibinfo {author} {\bibfnamefont {P.~R.}\ \bibnamefont {Berman}},\ and\
  \bibinfo {author} {\bibfnamefont {B.}~\bibnamefont {Dubetsky}},\ }\bibfield
  {title} {\bibinfo {title} {Time-domain de $\mathrm{B}$roglie wave
  interferometry},\ }\href {https://doi.org/10.1103/PhysRevLett.79.784}
  {\bibfield  {journal} {\bibinfo  {journal} {Phys. Rev. Lett.}\ }\textbf
  {\bibinfo {volume} {79}},\ \bibinfo {pages} {784} (\bibinfo {year}
  {1997})}\BibitemShut {NoStop}%
\bibitem [{\citenamefont {Barrett}\ \emph {et~al.}(2011)\citenamefont
  {Barrett}, \citenamefont {Chan}, \citenamefont {Mok}, \citenamefont {Carew},
  \citenamefont {Yavin}, \citenamefont {Kumarakrishnan}, \citenamefont {Cahn},\
  and\ \citenamefont {Sleator}}]{websiteAMOreview}%
  \BibitemOpen
  \bibfield  {author} {\bibinfo {author} {\bibfnamefont {B.}~\bibnamefont
  {Barrett}}, \bibinfo {author} {\bibfnamefont {I.}~\bibnamefont {Chan}},
  \bibinfo {author} {\bibfnamefont {C.}~\bibnamefont {Mok}}, \bibinfo {author}
  {\bibfnamefont {A.}~\bibnamefont {Carew}}, \bibinfo {author} {\bibfnamefont
  {I.}~\bibnamefont {Yavin}}, \bibinfo {author} {\bibfnamefont
  {A.}~\bibnamefont {Kumarakrishnan}}, \bibinfo {author} {\bibfnamefont
  {S.}~\bibnamefont {Cahn}},\ and\ \bibinfo {author} {\bibfnamefont
  {T.}~\bibnamefont {Sleator}},\ }\bibfield  {title} {\bibinfo {title} {Chapter
  3 - time-domain interferometry with laser-cooled atoms},\ }in\ \href
  {https://doi.org/https://doi.org/10.1016/B978-0-12-385508-4.00003-6} {\emph
  {\bibinfo {booktitle} {Advances in Atomic, Molecular, and Optical
  Physics}}},\ \bibinfo {series} {Advances In Atomic, Molecular, and Optical
  Physics}, Vol.~\bibinfo {volume} {60},\ \bibinfo {editor} {edited by\
  \bibinfo {editor} {\bibfnamefont {E.}~\bibnamefont {Arimondo}}, \bibinfo
  {editor} {\bibfnamefont {P.}~\bibnamefont {Berman}},\ and\ \bibinfo {editor}
  {\bibfnamefont {C.}~\bibnamefont {Lin}}}\ (\bibinfo  {publisher} {Academic
  Press},\ \bibinfo {year} {2011})\ pp.\ \bibinfo {pages} {119 --
  199}\BibitemShut {NoStop}%
\bibitem [{\citenamefont {Barrett}\ \emph {et~al.}(2016)\citenamefont
  {Barrett}, \citenamefont {Carew}, \citenamefont {Beica}, \citenamefont
  {Vorozcovs}, \citenamefont {Pouliot},\ and\ \citenamefont
  {Kumarakrishnan}}]{Atoms}%
  \BibitemOpen
  \bibfield  {author} {\bibinfo {author} {\bibfnamefont {B.}~\bibnamefont
  {Barrett}}, \bibinfo {author} {\bibfnamefont {A.}~\bibnamefont {Carew}},
  \bibinfo {author} {\bibfnamefont {H.~C.}\ \bibnamefont {Beica}}, \bibinfo
  {author} {\bibfnamefont {A.}~\bibnamefont {Vorozcovs}}, \bibinfo {author}
  {\bibfnamefont {A.}~\bibnamefont {Pouliot}},\ and\ \bibinfo {author}
  {\bibfnamefont {A.}~\bibnamefont {Kumarakrishnan}},\ }\bibfield  {title}
  {\bibinfo {title} {Prospects for precise measurements with echo atom
  interferometry},\ }\href {https://www.mdpi.com/2218-2004/4/3/19} {\bibfield
  {journal} {\bibinfo  {journal} {Atoms}\ }\textbf {\bibinfo {volume} {4}},\
  \bibinfo {pages} {19} (\bibinfo {year} {2016})}\BibitemShut {NoStop}%
\bibitem [{\citenamefont {Barrett}\ \emph {et~al.}(2013)\citenamefont
  {Barrett}, \citenamefont {Carew}, \citenamefont {Beattie},\ and\
  \citenamefont {Kumarakrishnan}}]{brynlePRA}%
  \BibitemOpen
  \bibfield  {author} {\bibinfo {author} {\bibfnamefont {B.}~\bibnamefont
  {Barrett}}, \bibinfo {author} {\bibfnamefont {A.}~\bibnamefont {Carew}},
  \bibinfo {author} {\bibfnamefont {S.}~\bibnamefont {Beattie}},\ and\ \bibinfo
  {author} {\bibfnamefont {A.}~\bibnamefont {Kumarakrishnan}},\ }\bibfield
  {title} {\bibinfo {title} {Measuring the atomic recoil frequency using a
  modified grating-echo atom interferometer},\ }\href
  {https://doi.org/10.1103/PhysRevA.87.033626} {\bibfield  {journal} {\bibinfo
  {journal} {Phys. Rev. A}\ }\textbf {\bibinfo {volume} {87}},\ \bibinfo
  {pages} {033626} (\bibinfo {year} {2013})}\BibitemShut {NoStop}%
\bibitem [{\citenamefont {Mok}\ \emph {et~al.}(2013)\citenamefont {Mok},
  \citenamefont {Barrett}, \citenamefont {Carew}, \citenamefont {Berthiaume},
  \citenamefont {Beattie},\ and\ \citenamefont {Kumarakrishnan}}]{carsonPRA}%
  \BibitemOpen
  \bibfield  {author} {\bibinfo {author} {\bibfnamefont {C.}~\bibnamefont
  {Mok}}, \bibinfo {author} {\bibfnamefont {B.}~\bibnamefont {Barrett}},
  \bibinfo {author} {\bibfnamefont {A.}~\bibnamefont {Carew}}, \bibinfo
  {author} {\bibfnamefont {R.}~\bibnamefont {Berthiaume}}, \bibinfo {author}
  {\bibfnamefont {S.}~\bibnamefont {Beattie}},\ and\ \bibinfo {author}
  {\bibfnamefont {A.}~\bibnamefont {Kumarakrishnan}},\ }\bibfield  {title}
  {\bibinfo {title} {Demonstration of improved sensitivity of echo
  interferometers to gravitational acceleration},\ }\href
  {https://doi.org/10.1103/PhysRevA.88.023614} {\bibfield  {journal} {\bibinfo
  {journal} {Phys. Rev. A}\ }\textbf {\bibinfo {volume} {88}},\ \bibinfo
  {pages} {023614} (\bibinfo {year} {2013})}\BibitemShut {NoStop}%
\bibitem [{\citenamefont {Janicke}\ and\ \citenamefont
  {Wilkens}(1994)}]{channeling}%
  \BibitemOpen
  \bibfield  {author} {\bibinfo {author} {\bibfnamefont {U.}~\bibnamefont
  {Janicke}}\ and\ \bibinfo {author} {\bibfnamefont {M.}~\bibnamefont
  {Wilkens}},\ }\bibfield  {title} {\bibinfo {title} {Atomic motion in a
  magneto-optical field},\ }\href {https://doi.org/10.1103/PhysRevA.50.3265}
  {\bibfield  {journal} {\bibinfo  {journal} {Phys. Rev. A}\ }\textbf {\bibinfo
  {volume} {50}},\ \bibinfo {pages} {3265} (\bibinfo {year}
  {1994})}\BibitemShut {NoStop}%
\bibitem [{\citenamefont {Romalis}\ \emph {et~al.}(1997)\citenamefont
  {Romalis}, \citenamefont {Miron},\ and\ \citenamefont
  {Cates}}]{RomalisBroad}%
  \BibitemOpen
  \bibfield  {author} {\bibinfo {author} {\bibfnamefont {M.~V.}\ \bibnamefont
  {Romalis}}, \bibinfo {author} {\bibfnamefont {E.}~\bibnamefont {Miron}},\
  and\ \bibinfo {author} {\bibfnamefont {G.~D.}\ \bibnamefont {Cates}},\
  }\bibfield  {title} {\bibinfo {title} {Pressure broadening of $\mathrm{Rb}$
  ${D}_{1}$ and ${D}_{2}$ lines by ${}^{3}\mathrm{He}$, ${}^{4}\mathrm{He}$,
  $\mathrm{N}_{2}$, and $\mathrm{Xe}$: Line cores and near wings},\ }\href
  {https://doi.org/10.1103/PhysRevA.56.4569} {\bibfield  {journal} {\bibinfo
  {journal} {Phys. Rev. A}\ }\textbf {\bibinfo {volume} {56}},\ \bibinfo
  {pages} {4569} (\bibinfo {year} {1997})}\BibitemShut {NoStop}%
\bibitem [{\citenamefont {Rotondaro}\ and\ \citenamefont
  {Perram}(1997)}]{CollBroad}%
  \BibitemOpen
  \bibfield  {author} {\bibinfo {author} {\bibfnamefont {M.~D.}\ \bibnamefont
  {Rotondaro}}\ and\ \bibinfo {author} {\bibfnamefont {G.~P.}\ \bibnamefont
  {Perram}},\ }\bibfield  {title} {\bibinfo {title} {Collisional broadening and
  shift of the rubidium $\mathrm{D}1$ and $\mathrm{D}2$ lines
  ($5^2\mathrm{S}_{1/2}\rightarrow 5^2\mathrm{P}_{1/2}, 5^2\mathrm{P}_{3/2}$)
  by rare gases, $\mathrm{H}_2$, $\mathrm{D}_2$, $\mathrm{N}_2$,
  $\mathrm{CH}_4$ and $\mathrm{CF}_4$},\ }\href
  {https://doi.org/https://doi.org/10.1016/S0022-4073(96)00147-1} {\bibfield
  {journal} {\bibinfo  {journal} {J. Quant. Spectrosc. Radiat. Transfer}\
  }\textbf {\bibinfo {volume} {57}},\ \bibinfo {pages} {497 } (\bibinfo {year}
  {1997})}\BibitemShut {NoStop}%
\bibitem [{\citenamefont {Corney}(1977)}]{Corney}%
  \BibitemOpen
  \bibfield  {author} {\bibinfo {author} {\bibfnamefont {A.}~\bibnamefont
  {Corney}},\ }\href {https://books.google.ca/books?id=KgVowQEACAAJ} {\emph
  {\bibinfo {title} {Atomic and Laser Spectroscopy}}},\ Oxford science
  publications\ (\bibinfo  {publisher} {Clarendon Press},\ \bibinfo {year}
  {1977})\BibitemShut {NoStop}%
\bibitem [{\citenamefont {Beica}\ \emph {et~al.}(2019)\citenamefont {Beica},
  \citenamefont {Pouliot}, \citenamefont {Carew}, \citenamefont {Vorozcovs},
  \citenamefont {Afkhami-Jeddi}, \citenamefont {Vacheresse}, \citenamefont
  {Carlse}, \citenamefont {Dowling}, \citenamefont {Barron},\ and\
  \citenamefont {Kumarakrishnan}}]{HerminaRSI}%
  \BibitemOpen
  \bibfield  {author} {\bibinfo {author} {\bibfnamefont {H.~C.}\ \bibnamefont
  {Beica}}, \bibinfo {author} {\bibfnamefont {A.}~\bibnamefont {Pouliot}},
  \bibinfo {author} {\bibfnamefont {A.}~\bibnamefont {Carew}}, \bibinfo
  {author} {\bibfnamefont {A.}~\bibnamefont {Vorozcovs}}, \bibinfo {author}
  {\bibfnamefont {N.}~\bibnamefont {Afkhami-Jeddi}}, \bibinfo {author}
  {\bibfnamefont {T.}~\bibnamefont {Vacheresse}}, \bibinfo {author}
  {\bibfnamefont {G.}~\bibnamefont {Carlse}}, \bibinfo {author} {\bibfnamefont
  {P.}~\bibnamefont {Dowling}}, \bibinfo {author} {\bibfnamefont
  {B.}~\bibnamefont {Barron}},\ and\ \bibinfo {author} {\bibfnamefont
  {A.}~\bibnamefont {Kumarakrishnan}},\ }\bibfield  {title} {\bibinfo {title}
  {Characterization and applications of auto-locked vacuum-sealed diode lasers
  for precision metrology},\ }\href {https://doi.org/10.1063/1.5112760}
  {\bibfield  {journal} {\bibinfo  {journal} {Rev. Sci. Inst.}\ }\textbf
  {\bibinfo {volume} {90}},\ \bibinfo {pages} {085113} (\bibinfo {year}
  {2019})}\BibitemShut {NoStop}%
\bibitem [{\citenamefont {Pouliot}\ \emph
  {et~al.}(2018{\natexlab{b}})\citenamefont {Pouliot}, \citenamefont {Beica},
  \citenamefont {Carew}, \citenamefont {Vorozcovs}, \citenamefont {Carlse},\
  and\ \citenamefont {Kumarakrishnan}}]{SPIEPWest}%
  \BibitemOpen
  \bibfield  {author} {\bibinfo {author} {\bibfnamefont {A.}~\bibnamefont
  {Pouliot}}, \bibinfo {author} {\bibfnamefont {H.~C.}\ \bibnamefont {Beica}},
  \bibinfo {author} {\bibfnamefont {A.}~\bibnamefont {Carew}}, \bibinfo
  {author} {\bibfnamefont {A.}~\bibnamefont {Vorozcovs}}, \bibinfo {author}
  {\bibfnamefont {G.}~\bibnamefont {Carlse}},\ and\ \bibinfo {author}
  {\bibfnamefont {A.}~\bibnamefont {Kumarakrishnan}},\ }\bibfield  {title}
  {\bibinfo {title} {{Auto-locking waveguide amplifier system for lidar and
  magnetometric applications}},\ }in\ \href
  {https://doi.org/10.1117/12.2286952} {\emph {\bibinfo {booktitle} {High-Power
  Diode Laser Technology XVI}}},\ Vol.\ \bibinfo {volume} {10514},\ \bibinfo
  {editor} {edited by\ \bibinfo {editor} {\bibfnamefont {M.~S.}\ \bibnamefont
  {Zediker}}},\ \bibinfo {organization} {International Society for Optics and
  Photonics}\ (\bibinfo  {publisher} {SPIE},\ \bibinfo {year} {2018})\ pp.\
  \bibinfo {pages} {152 -- 159}\BibitemShut {NoStop}%
\bibitem [{sta(2015)}]{stanfordResearch}%
  \BibitemOpen
  \href@noop {} {\emph {\bibinfo {title} {Model PRS10 Rubidium Frequency
  Standard}}},\ \bibinfo {organization} {Stanford Research Systems} (\bibinfo
  {year} {2015})\BibitemShut {NoStop}%
\bibitem [{\citenamefont {Kluttz}\ \emph {et~al.}(2013)\citenamefont {Kluttz},
  \citenamefont {Averett},\ and\ \citenamefont {Wolin}}]{Broad2013}%
  \BibitemOpen
  \bibfield  {author} {\bibinfo {author} {\bibfnamefont {K.~A.}\ \bibnamefont
  {Kluttz}}, \bibinfo {author} {\bibfnamefont {T.~D.}\ \bibnamefont
  {Averett}},\ and\ \bibinfo {author} {\bibfnamefont {B.~A.}\ \bibnamefont
  {Wolin}},\ }\bibfield  {title} {\bibinfo {title} {Pressure broadening and
  frequency shift of the ${D}_{1}$ and ${D}_{2}$ lines of $\mathrm{Rb}$ and
  $\mathrm{K}$ in the presence of ${}^{3}\mathrm{He}$ and $\mathrm{N}_{2}$},\
  }\href {https://doi.org/10.1103/PhysRevA.87.032516} {\bibfield  {journal}
  {\bibinfo  {journal} {Phys. Rev. A}\ }\textbf {\bibinfo {volume} {87}},\
  \bibinfo {pages} {032516} (\bibinfo {year} {2013})}\BibitemShut {NoStop}%
\bibitem [{\citenamefont {Lide}(2007)}]{RbVP}%
  \BibitemOpen
  \bibinfo {editor} {\bibfnamefont {D.~R.}\ \bibnamefont {Lide}},\ ed.,\
  \href@noop {} {\emph {\bibinfo {title} {CRC Handbook of Chemistry and
  Physics, 88th Edition}}}\ (\bibinfo  {publisher} {CRC Press},\ \bibinfo
  {year} {2007})\BibitemShut {NoStop}%
\bibitem [{\citenamefont {Steck}(2019)}]{Steck87}%
  \BibitemOpen
  \bibfield  {author} {\bibinfo {author} {\bibfnamefont {D.}~\bibnamefont
  {Steck}},\ }\bibfield  {title} {\bibinfo {title} {Rubidium 87 $\mathrm{D}$
  line data (version 2.2.1)},\ }\href {https://steck.us/alkalidata/} {\
  (\bibinfo {year} {2019})}\BibitemShut {NoStop}%
\bibitem [{\citenamefont {Klein}\ and\ \citenamefont {Smith}(1968)}]{NIST}%
  \BibitemOpen
  \bibfield  {author} {\bibinfo {author} {\bibfnamefont {M.}~\bibnamefont
  {Klein}}\ and\ \bibinfo {author} {\bibfnamefont {F.}~\bibnamefont {Smith}},\
  }\bibfield  {title} {\bibinfo {title} {Tables of collision integrals for the
  (m,6) potential function for 10 values of m},\ }\href@noop {} {\bibfield
  {journal} {\bibinfo  {journal} {J. Res. Natl. Bur. Stand. Sec. A}\ }\textbf
  {\bibinfo {volume} {72A}},\ \bibinfo {pages} {359} (\bibinfo {year}
  {1968})}\BibitemShut {NoStop}%
\end{thebibliography}

\providecommand{\noopsort}[1]{}\providecommand{\singleletter}[1]{#1}%

\end{document}